\documentclass[]{aastex631}
\usepackage{amsmath}
\usepackage{multirow}
\usepackage{url}
\usepackage{ulem}
\received{2025 September 22}
\revised{2026 March 22}
\accepted{2026 March 24}

\submitjournal{ApJ}

\shorttitle{Protoplanetary disk evolution in a low-metallicity environment}

\begin{document}


\title{Protoplanetary Disk Evolution in a Low-Metallicity Environment:
  JWST's First Mid-Infrared Census of Low-Mass Stars}

\correspondingauthor{Chikako Yasui}
\email{ck.yasui@gmail.com}

\author[0000-0003-3579-7454]{Chikako Yasui}
\affiliation{National Astronomical Observatory of Japan, 2-21-1 Osawa, Mitaka, Tokyo 181-8588, Japan}
\affiliation{Department of Astronomical Science, School of Physical
  Science, SOKENDAI (The Graduate University for Advanced Studies),
  2-21-1 Osawa, Mitaka, Tokyo 181-8588, Japan}

\author[0000-0003-1604-9127]{Natsuko Izumi}
\affiliation{National Astronomical Observatory of Japan, 2-21-1 Osawa, Mitaka, Tokyo 181-8588, Japan}
\affiliation{Graduate School of Science, Nagoya University, Furo-cho, Chikusa-ku, Nagoya, Aichi 464-8602, Japan}


\author[0000-0003-0769-8627]{Masao Saito}
\affiliation{National Astronomical Observatory of Japan, 2-21-1 Osawa, Mitaka, Tokyo 181-8588, Japan}
\affiliation{Department of Astronomical Science, School of Physical
  Science, SOKENDAI (The Graduate University for Advanced Studies),
  2-21-1 Osawa, Mitaka, Tokyo 181-8588, Japan}

\author[0000-0003-0778-0321]{Ryan M. Lau}
\affiliation{IPAC, Mail Code 100-22, Caltech, 1200 E. California Blvd., Pasadena, CA 91125, USA}

\author[0000-0003-4578-2619]{Naoto Kobayashi}
\affiliation{Institute of Astronomy, School of Science, University of Tokyo, 2–21–1 Osawa, Mitaka, Tokyo 181–0015, Japan}
\affiliation{Kiso Observatory, Institute of Astronomy, School of Science, University of Tokyo, 10762–30 Mitake, Kiso-machi, Kiso-gun, Nagano 397–0101, Japan}

\author[0000-0001-5644-8830]{Michael E. Ressler}
\affiliation{Jet Propulsion Laboratory, California Institute of Technology, 4800 Oak Grove Drive, Pasadena, CA 91109, USA}




\begin{abstract}

This study presents the first high-resolution, high-sensitivity
mid-infrared (MIR) investigation of protoplanetary disks in a
low-metallicity environment, using JWST/NIRCam and MIRI observations
of Digel Cloud 2, a star-forming region in the outer Galaxy ($D \simeq
8$ kpc, ${\rm [M/H]} \simeq -0.7$ dex). It hosts two very young
($\sim$0.1 Myr) embedded clusters, Cloud 2-N and Cloud 2-S, offering a
window into disk evolution under conditions analogous to the early
universe, where low metallicity implies reduced dust content. Imaging
across 1--20 $\mu$m, including F770W and complementary bands (F356W,
F444W, F405N), enables probing disk properties with unprecedented
spatial resolution and stellar mass sensitivity down to $\sim$0.1
$M_\odot$.
Among 89 and 95 sources detected in F770W in Cloud 2-N and 2-S,
respectively, we identify candidate stellar-mass cluster members using
infrared photometry, from which stellar mass and extinction are
estimated. Among these, $\simeq$75 \% retain optically thick disks in
both clusters based on MIR SED slopes, consistent with similarly aged
solar-metallicity regions. In contrast, a lack of 2 $\mu$m excess
suggests diminished inner disk emission, possibly due to enhanced
silicate grains with low sublimation temperatures.
Using the F405N narrow-band filter covering Br$\alpha$, we detect
accretion signatures in $\simeq$35 \% of sources selected by
extinction criteria, with rates
$\gtrsim$10$^{-6}$\,$M_\odot$\,yr$^{-1}$, comparable to or exceeding
those in nearby low-mass stars. Brown dwarf candidates, identified
across multiple bands including F770W and shorter wavelengths, exhibit
a high disk fraction of $\sim$75 \%, indicating robust disk retention
across mass ranges even under low-metallicity conditions.

 \end{abstract}



\section{Introduction} \label{sec:Intro}

Protoplanetary disks, composed of gas and dust, are the birthplaces of
planets. Among these components, dust plays a crucial role in the
formation of planetary cores. {In recent years, a wide variety of
  exoplanets have been discovered \citep{Valencia2025}}, and a clear
correlation has been established between the presence of planets and
the metallicity of their host stars. This correlation, known as the
``planet–metallicity relation'' \citep{Fischer2005}, suggests the
importance of metallicity in the planet formation process. {However,
most of these studies have focused on environments with relatively
high metallicity specifically in the solar neighborhood
\citep[e.g.,][]{Williams2016}}, leaving our
understanding of disk formation and evolution in the early universe or
in low-metallicity environments still incomplete. Investigating the
properties of disks in low-metallicity environments is essential for
understanding planet formation in the early universe and for testing
the universality of planet formation theories. In particular,
metallicity is thought to play a significant role in disk evolution
and dissipation processes, and observational verification of this
effect is highly desired.

To elucidate the dissipation processes of protoplanetary disks,
extensive observational efforts have been made using advanced
instruments such as the Hubble Space Telescope, the Spitzer Space
Telescope, large ground-based telescopes, and ALMA. {Recent
  developments have enabled the direct detection and spatial
  resolution of nearby disks (e.g., \citealt{Andrews2018})}, revealing
detailed structures. However, capturing the actual dissipation process
remains challenging even with current technology. As a result,
{various theoretical models have been proposed \citep{Alexander2014}}
based on fragmented observational data, but testing these models
remains difficult. In this study, we focus on metallicity—a physical
parameter believed to be closely related to disk dissipation—and aim
to evaluate its role observationally. Metallicity, defined as the
abundance of elements heavier than hydrogen and helium, is a
fundamental factor in astrophysical processes, influencing physical
processes (e.g., heating and cooling) and chemical mechanisms that
govern disk evolution. Although the metal content in protoplanetary
disks is only a few percent, it directly contributes to the formation
of planetary cores, making metallicity a key element in planet
formation.

To investigate the evolution of protoplanetary disks in
low-metallicity environments, {we studied Digel Cloud 2 (hereafter
  referred to as Cloud 2)}, a star-forming region located in the outer
part of the Galaxy, where the metallicity is approximately one-tenth
that of the Sun.
Near-infrared (NIR) observations ($\lambda \le 2$ $\mu$m) conducted
with the Subaru Telescope enabled the detection of objects down to
$\sim$0.1 $M_\odot$ and provided {observational} evidence
suggesting that disk lifetime is significantly shorter ($\sim$1 Myr)
than the typical values observed in the solar neighborhood
($\sim$5--10 Myr) \citep{{Yasui2009}, {Yasui2010}, {Yasui2016a},
  {Yasui2016b}, {Yasui2021}}.
{This observational evidence led to the hypothesis that, if
  protoplanetary disks have shorter lifetimes in low-metallicity
  environments, planet formation may be suppressed, resulting in a
  lower frequency of planets.
  This idea offers a possible explanation for the observed
  planet–metallicity relation.}
This trend has also been supported by statistical analyses using
archival data \citep{Patra2024}. However, NIR observations are limited
to the hot inner regions of disks ($\sim$0.1 au, $\sim$1500 K),
requiring many assumptions to discuss overall disk evolution.
Longer-wavelength observations have been attempted to probe disk
regions beyond those accessible by the NIR observations. For example,
Spitzer/IRAC observations up to 4.5 $\mu$m enable a more reliable
assessment of disk presence, although they still do not sufficiently
trace the outer disk structures. In relatively nearby regions within
the outer Galaxy, where the metallicity is significantly lower than
solar values, these observations are generally sensitive only to
intermediate-mass stars, while low-mass stars remain largely
undetected \citep{Yasui2021_Spitzer}.

Among low-metallicity regions within the Milky Way, Sh 2-284 has been
the subject of multi-wavelength studies. Although its metallicity is
not very low ($-$0.3 dex),
{it is often used to investigate disk evolution in low-metallicity
  environments. Its age is estimated to be young, typically in the
  range of 1--3 Myr based on photometric and spectroscopic analyses
  \citep[e.g.,][]{{Cusano2011}, {Kalari2015}, {Guarcello2021}}.}
Several works have reported contrasting results: \citet{Puga2009} and
\citet{Cusano2011} support dust disk fractions similar to those in the
solar neighborhood, while \citet{Guarcello2021} report a lower disk
fraction than expected for its age. Measurements of mass accretion
rates using optical H$\alpha$ spectroscopy suggest values comparable
to solar-metallicity objects \citep{Kalari2015}. However, these
observations are generally limited to objects up to $\sim$1 $M_\odot$,
and thus cannot rule out mass-dependent disk evolution (e.g.,
\citealt{Ribas2014}).

In extragalactic low-metallicity environments such as the nearby dwarf
irregular galaxies LMC and SMC, H$\alpha$ imaging with HST has
revealed significantly higher mass accretion rates compared to
solar-neighborhood stars, suggesting that gas disk lifetimes may be
extremely long (\citealt{De Marchi2010} and references therein). This
trend has been further supported by spectroscopic observations with
JWST/NIRSpec, which detected strong H$\alpha$ emission \citep{De
  Marchi2024}. Additionally, JWST observations with NIRCam and MIRI
have enabled the detection of dust disks around stars down to $\sim$1
$M_\odot$ \citep{{Jones2023},{Habel2024}}.  Nevertheless, direct
comparisons with low-mass stars in the solar neighborhood remain
challenging. To explore disk evolution in low-metallicity
environments, it is essential to detect objects down to $\sim$0.1
$M_\odot$ in the MIR, as has been achieved in previous
studies of the solar neighborhood.

{Ground-based telescopes face significant limitations in this
  wavelength range due to thermal emission and molecular absorption
  from Earth's atmosphere, particularly beyond 3 $\mu$m. The Spitzer
  Space Telescope, while free from atmospheric interference, had a
  small aperture and insufficient spatial resolution, making it
  difficult to resolve individual sources. These limitations highlight
  the need for space-based observations at longer wavelengths to
  investigate faint and low-mass objects in star-forming regions.}

{The James Webb Space Telescope (JWST) overcomes these challenges,
  offering high spatial resolution and sensitivity across a wide
  wavelength range. In particular, MIRI enables detection of low-mass
  objects ($\sim$0.1 $M_\odot$) in the 7.7\,$\mu$m band, providing new
  insights into regions previously inaccessible. Additionally, JWST's
  F405N narrowband filter allows the evaluation of mass accretion
  rates through hydrogen emission lines, contributing to our
  understanding of gas disk evolution.}

In this study, we conducted {high-sensitivity} and high-resolution
MIR ($\lambda \le 20$ $\mu$m) observations using JWST {to
  investigate the evolution of protoplanetary disks in low-metallicity
  environments. Our targets are two star-forming clusters,} Cloud 2-N
and Cloud 2-S, located in Cloud 2—a low-metallicity {($-$0.7
  dex)} star-forming region situated in the outermost part of the
Galaxy ($R_G \gtrsim 15$ kpc; \citealt{Izumi2024}).
Based on the Gaia DR3 parallax {for source ID 461019899768797824
  (0.0992$\pm$0.0193 mas; \citealt{Gaia2023}), the corresponding
  geometric distance estimate from \citet{Bailer-Jones2021} is
  $7.9^{+1.2}_{-1.1}$ kpc.}
Both Cloud 2-N and Cloud 2-S are considered to be nearly coeval with
an estimated age of $\simeq$0.1 Myr \citep{Yasui2024}{, similar to
  some of the youngest embedded clusters reported in the solar
  neighborhood, such as $\rho$ Ophiuchi, NGC 2024, and NGC 1333 (all
  younger than 1 Myr; e.g., \citealt{Lada2003},
  \citealt{Wilking2008}).}

{To investigate disk evolution in detail, this study primarily
  utilizes data from NIRCam's longer wavelength channel and MIRI,
  which are particularly effective for probing disk morphology,
  lifetimes, and dust properties. This approach enables an extended
  evaluation of the impact of metallicity on disk evolution,
  especially for low-mass stars that were previously difficult to
  observe.}

{Previous and ongoing JWST studies of Cloud 2} suggest that star
formation in low-metallicity environments differs significantly from
that in the solar neighborhood.
{Specifically, \citep{Izumi2024} reported} strong jets and outflows were
detected, extending in various directions, contrary to the previously
suggested picture of relatively quiescent star formation.
Furthermore, {\citet{Yasui2024}, using NIRCam's short-wavelength
  channel, derived an initial mass function (IMF) for the young
  clusters with a peak mass of $\sim$0.03 $M_\odot$,}
nearly an order of magnitude lower than the typical value ($\sim$0.3
$M_\odot$) observed in solar-metallicity environments.
These findings imply that the physical conditions of star formation in
low-metallicity environments may be fundamentally different,
potentially influencing disk evolution as well.
{The present study complements these works by focusing on longer
  wavelengths (NIRCam long-wavelength channel and MIRI).}

{In this paper, we investigate} the evolution of protoplanetary
disks in low-metallicity environments based on JWST observations of
the Cloud 2-N and Cloud 2-S clusters.
The paper is organized as follows. Section~\ref{sec:Obs} describes the
observational overview, data reduction procedures, and photometry. In
Section~\ref{sec:Result}, we present the observational results:
Section~\ref{sec:cm} discusses the mass detection limits derived from
JWST color-magnitude diagrams; Section~\ref{sec:sed} classifies disks
based on the slope of the MIR spectral energy distribution
(SED);
Section~\ref{sec:ident_cluster} identifies possible cluster members in
Cloud 2-N and Cloud 2-S using an extinction-limited sample
($A_V$-limited sample); Section~\ref{sec:dust_diskfraction} evaluates
the optically thick disk fraction of stellar-mass objects based on a
mass- and extinction-limited sample (Mass-$A_V$-limited sample).
Section~\ref{sec:discussion} provides a discussion based on the
obtained results. Specifically, Section~\ref{sec:dust_disk} addresses
the evolution of dust disks in the Cloud 2 clusters;
Section~\ref{sec:gas_disk} explores implications for the evolution of
gas disks; Section~\ref{sec:other_prop} investigates other properties
of disk-classified objects; Section~\ref{sec:diskevolve_lowmeta}
discusses the characteristics of disk evolution in low-metallicity
environments; and Section~\ref{sec:disk_bd} examines the disk
occurrence rate of brown dwarfs based on JWST/NIRCam data. Finally,
Section~\ref{sec:conclusion} summarizes the conclusions of this study.

\section{Observations, Data Reduction, and Photometry} \label{sec:Obs}

\subsection{JWST/NIRCam and MIRI Imaging and Data Reduction} \label{sec:imaging}

As a part of the GTO of the JWST/MIRI instrument team (GTO 1237; PI:
Michael Ressler), we obtained images of Digel Cloud 2 from 1 to 20
$\mu$m (F115W/F150W/F200W/F356W/F444W/F405N-band data with NIRCam, and
F770W/F1280W/F2100W-band data with MIRI) on 2023 January 17 (UT).
The observations of Cloud 2-N NIRCam and Cloud 2-S MIRI were carried
out individually, and
NIRCam observations of Cloud 2-S and MIRI observations of Cloud 2-N
were performed simultaneously {as} NIRCam-MIRI parallel
observations because the separation of the two instruments' fields of
view matched the spatial distance between the Cloud 2 regions.
All of these observations were carried out within 24 hours, taking
into account the possible variabilities of YSO luminosities and
background levels.

{The observational setup was designed to cover the Cloud 2-N and
  Cloud 2-S regions using both NIRCam and MIRI, with a combination of
  individual and parallel observations. The full observational
  configuration, including pointing strategy and field coverage, is
  described in detail in \citet{Izumi2024}, which provides a
  comprehensive overview of the JWST program.}

The pixel scale and field of view of MIRI are 0$\farcs$11 and
$74\arcsec \times 113\arcsec$, respectively.
The MIRI maps obtained for Cloud 2-N have a center of approximately
$\alpha = 2^{\rm h} 48^{\rm m} 44\fs7,
\delta = +58\arcdeg 29\arcmin 29\arcsec0$ (J2000.0),
encompassing approximately two fields of view with a position angle
(PA) of $\simeq$90 deg. 
The coordinates for Cloud 2-S are approximately
$\alpha = 2^{\rm h} 48^{\rm m} 30\fs7,
\delta = +58\arcdeg 23\arcmin 40\arcsec2$ (J2000.0),
with a field of view encompassing approximately one field of view and
a PA of $\simeq$90 deg.
For the NIRCam observations, only module B was used, and 
the pixel scale and field of view of short/long channels are
0$\farcs$031/0$\farcs$063 and $2\farcm2 \times 2\farcm2$, respectively.
For Cloud 2-N, the center coordinates are
$\alpha = 2^{\rm h} 48^{\rm m} 46\fs1,
\delta = +58\arcdeg 29\arcmin 31\arcsec9$ (J2000.0),
with images covering $2\farcm2 \times 2\farcm2$.
For Cloud 2-S, the center coordinates are
$\alpha = 2^{\rm h} 48^{\rm m} 32\fs3,
\delta = +58\arcdeg 23\arcmin 19\arcsec9$ (J2000.0),
with images covering $3\farcm1 \times 2\farcm4$.
As discussed in \citet{Izumi2024}, observations were planned to cover
target star-forming clusters here, Digel Cloud 2 clusters, in all
bands.
For this purpose, the gaps on the NIRCam short channels were designed
so that they did not fall within the area of the clusters.

Data calibration was performed with JWST Science Calibration pipeline
version number 1.16.1 \citep{Bushouse2024}, the Calibration Reference
Data System version 11.18.4, and CRDC context \verb|jwst_1293.pmap|.
The images were then combined into one mosaic at each of the nine bands. 
The same custom parameters as in \citet{Izumi2024} were used for
calibration, except for ``\verb|abs_refcat|'' in the ``tweakreg'' step
of the Image3 pipeline, which used GAIADR3, where GAIADR3 was selected
instead of GAIADR2, as in \citet{Izumi2024}.
Figure~\ref{fig:CL2NScl} shows pseudocolor images created from the
Stage 3 combined images using the F356W (blue), F444W (green), and
F700W (red) bands. The images present zoomed-in views of the Cloud 2-N
and Cloud 2-S clusters.

\subsection{JWST/NIRCam and MIRI Source Detection and Photometry}
\label{sec:phot}

The detection and photometry of objects in the obtained NIRCam and
MIRI data were performed using STARBUGII \citep{Nally2023}. STARBUGII is a
software optimized for JWST data, designed for extracting point
sources, performing photometry, and integrating photometric results in
highly crowded and complex regions. 
It utilizes PHOTUTILS \citep{Bradley2022} and astropy.
The detection algorithm identifies sources that exceed a 5$\sigma$
threshold above the locally estimated sky background and applies
morphological criteria based on sharpness and roundness. These
criteria are effective in filtering out cosmic rays and excluding
spatially resolved background galaxies from the source list.

For NIRCam F356W/F444W/F405N and MIRI F770W, object detection and
aperture photometry were performed on Stage 2 data. The basic
methodology followed \citet{Habel2024}, which dealt with star-forming
regions similar to the targets here.
The detection and photometry parameters used in STARBUGII were
primarily the same as those in \citet{Habel2024}. For the F356W and
F405N bands, which were not used in \citet{Habel2024}, the same
parameters as those for F444W in that study were applied.

For the matching parameters for NIRCam long channel data,
\verb|MATCH_THRESH| was set to the same value as in \citet{Habel2024}
for both Cloud 2-N and Cloud 2-S. Since the Cloud 2-N images were
obtained with four dithers, we followed \citet{Habel2024} and included
only objects detected in at least three dithers (\verb|NEXP_THRESH=3|)
in the catalog.
In contrast, for Cloud 2-S, the cluster region was covered by two sets
of four dithers each, and objects needed to be detected in at least 5
dithers (\verb|NEXP_THRESH=5|) to be included in the catalog.
These matching thresholds were chosen to minimize the inclusion of
transient or spurious detections, such as those caused by cosmic rays
or detector artifacts, and to enhance the reliability of identifying
true astrophysical sources across multiple exposures.

By first detecting objects in the Stage 2 images and then matching
them across multiple frames for the NIR long channel data, we aim to
prevent the misidentification of cosmic rays, detector artifacts, or
other transient sources as real astronomical objects.
For Cloud 2-S, increasing the detection threshold further might
improve the accuracy of source identification. However, the primary
objective of this study is to classify disks based on photometric
measurements in the F356W, F444W, and F770W bands. Even if a false
detection occurs in one band, it will not affect the final statistical
analysis unless the same misidentification happens in all three
bands. Therefore, we prioritized maximizing the number of sources
included in the photometry over minimizing the false detection rate.

For MIRI F770W, source detection and aperture photometry were also
performed using Stage 2 data. For both Cloud 2-N and 2-S, the cluster
region was covered with a single set of four frame dithers, and a
detection threshold of \verb|NEXP_THRESH=3| was adopted.

For aperture correction, the value for 80 \% encircled energy
presented in \verb|jwst_nircam_apcorr_0004.fits| from the Calibration
Reference Data System (CRDS) was adopted, while for MIRI, the 80\%
encircled energy from \verb|jwst_miri_apcorr_0010.fits| in CRDS was
adopted.
AB magnitudes were converted to Vega magnitudes using the CRDS files
\verb|jwst_nircam_abvegaoffset_0001.asdf| and
\verb|jwst_miri_abvegaoffset_0001.asdf|.

Among the bands primarily analyzed in this paper (F356W, F444W, F405N,
and F770W), F356W has the highest sensitivity and spatial
resolution. Therefore, matching is performed using the objects
detected in this band as a reference.
Before matching, each source's photometric uncertainty is evaluated,
and sources with a magnitude error exceeding 10 percent or data flags
indicating poor quality are removed from the individual catalogs.
A separation threshold of $0\farcs1$ was used for matching F356W with
the other bands.

The 10$\sigma$ detection limits for the primary bands analyzed in this
study are as follows: $m_{\rm F356W} = 22.3$ mag, $m_{\rm F444W} =
21.8$ mag, $m_{\rm F405W} = 18.9$ mag, and $m_{\rm F777W} = 19.5$ mag.
These values are used to derive the mass detection limits from the
color--magnitude diagrams (see Section~\ref{sec:cm}).
As a result a total of 126, 123, 69, and 89 objects in the Cloud 2-N
cluster region were detected in F356W, F444W, F405N, and F770W,
respectively, while {164, 158, 85, and 95 objects} in the Cloud 2-S
cluster region were detected in the corresponding bands.
For the NIRCam short channel data (F115W, F150W, and F200W), the
photometric results from \citet{Yasui2024} were adopted.

\subsection{JWST/NIRCam and MIRI Sensitivity and Spatial Resolution}

The mass detection limits for this observation are summarized in
\citet{Izumi2024}; however, caution is advised when interpreting the
exact values due to differences in the adopted distance to Cloud 2 and
slight variations in photometric methods. The wide band filters of
NIRCam (F115W, F150W, F200W, F356W, and F444W) reach a mass detection
limit of approximately 0.01 $M_\odot$, while the narrow band filter
(F405N) has a limit of $\simeq$0.1 $M_\odot$.
In contrast, the mass detection limit for the MIRI F770W filter is
also $\simeq$0.1 $M_\odot$, {based on 10$\sigma$ limiting
  magnitudes of 19.1--19.2 mag (Vega). The corresponding mass
  detection limits for the F1280W and F2100W filters are $\simeq$1
  $M_\odot$ and $>$1.4 $M_\odot$, based on 10$\sigma$ limiting
  magnitudes of 17.1--17.3 mag and 14.2--14.3 mag (Vega),
  respectively—more than an order of magnitude higher.}
\citet{Izumi2024} assumes a distance of $D = 12$ kpc to Cloud 2,
whereas in this study, we adopt a more likely {and smaller} value
of $D = 7.9$\,kpc.
This change reduces the distance modulus from 15.4 to 14.5 mag (a
difference of 0.9 mag), which lowers the mass detection limits
compared to \citet{Izumi2024} (see Section~\ref{sec:cm} for
quantitative details). While these changes affect the absolute values,
we confirm in the following section that the overall conclusions
regarding the presence and distribution of low-mass sources remain
qualitatively unchanged.

Regarding spatial resolution, it is important to note that the full
width at half maximum (FWHM) is generally proportional to the observed
wavelength. Specifically, the FWHM is $<$$0\farcs1$ for NIRCam's
short-wavelength channel, $<$$0\farcs2$ for its long-wavelength
channel, $\lesssim$$0\farcs3$ for MIRI F770W, and $\simeq$$0\farcs7$
for MIRI F2100W.

These spatial resolution and sensitivity characteristics are critical
for the scientific goals of this study, particularly the investigation
of disk properties in low-metallicity environments.
Specifically, using data at wavelengths longer than 4 $\mu$m
allows the detection of not only optically thick disks but also
optically thin disks \citep{Sicilia-Aguilar2006}. However, in the
MIR wavelength range, in particular, sensitivity and spatial
resolution decrease toward longer wavelengths. Considering both
sensitivity and spatial resolution, this study primarily utilizes data
up to the F770W band to fully leverage JWST's high sensitivity (mass
detection limit of down to $\simeq$0.1 $M_\odot$) and spatial
resolution ($\lesssim$0\farcs3). A comprehensive analysis of the
entire star-forming region using data up to F2100W will be presented
in a future paper.

The NIRCam long-wavelength channel bands F356W and F444W, and the MIRI
band F770W, roughly correspond to Spitzer/IRAC bands 1 (3.6 $\mu$m), 2
(4.5 $\mu$m), and 4 (8.0 $\mu$m), respectively. Although MIRI also has
an F560W band that roughly corresponds to Spitzer band 3 (5.8 $\mu$m),
data for this band were not obtained due to time constraints.
While Spitzer/IRAC observations of nearby star-forming regions ($D
\simeq 500$ pc) have FWHM values ranging from 1\farcs6 to 1\farcs9 in
the 3.6--8.0 $\mu$m range, JWST observations at $D \simeq 8$ kpc
achieve ${\rm FWHM} \simeq 0\farcs2$--0$\farcs3$, resulting in
comparable physical resolution.
In terms of sensitivity, the limiting magnitudes for Spitzer/IRAC
bands 1, 2, and 4 are approximately 18, 17.5, and 16.5 mag,
respectively \citep[e.g.,][]{Williams2016}. To match these
sensitivities at a distance of 8 kpc, a detection limit approximately
$\Delta m = 6$ mag fainter is required, which is achieved in the
present JWST observations.

These capabilities enable, for the first time, a direct comparison
between the outermost regions of the Milky Way and nearby star-forming
regions, in terms of both spatial resolution and sensitivity. This
allows us to investigate disk populations in environments previously
inaccessible to infrared studies, and to assess whether disk lifetimes
and properties differ significantly in low-metallicity outer Galaxy
regions.

\section{Results} \label{sec:Result}

\subsection{JWST Color--magnitude Diagrams: Mass Detection Limits}
\label{sec:cm}

To evaluate the mass detection limits relevant to disk classification,
we analyzed color--magnitude diagrams using the NIRCam F356W/F444W
bands and the MIRI F770W band. These bands are sensitive to disk
excess emission and allow for distinguishing between optically thick
and thin disks \citep{Sicilia-Aguilar2006}.

Figure~\ref{fig:CM_MIRI_NIRCam} shows the MIRI-NIRCam color--magnitude
diagrams, $(m_{\mathrm{F356W}} - m_{\mathrm{F770W}})$
vs. $m_{\mathrm{F770W}}$ diagrams for Cloud 2-N and Cloud 2-S. We
selected sources within the cluster regions identified by
\citet{Yasui2024}, shown as green ellipses in
Figure~\ref{fig:CL2NScl}, and included only {objects} detected at
$\geq$10$\sigma$ in all three bands.
The sensitivity limits, corresponding to the 10$\sigma$ limiting
magnitudes of the F356W and F770W bands, are indicated by dashed
lines.

The age of the Cloud 2 clusters is more likely to be $\simeq$0.1 Myr
\citep{Kobayashi2008, Yasui2024}.
To represent such young, low-metallicity populations, we adopted the
0.1 Myr isochrone from \citet{{D'Antona1997}, {D'Antona1998}}, which
is the only available model that simultaneously covers this very young
age and low-mass range (0.017--3\,$M_\odot$). For stellar atmospheres,
we used BT-NextGen models \citep{Allard1997}, which are valid for
$T_{\rm eff} \ge 2700$\,K, and BT-Settl models \citep{Allard2003},
which are valid for $T_{\rm eff} < 2800$\,K, connecting the two grids
at 0.1\,$M_\odot$ to ensure continuity across the mass range. Both
sets of models were adopted at solar metallicity (0 dex), because
BT-Settl provides only limited coverage for subsolar metallicities,
where models exist only for $T_{\rm eff} \ge 2600$\,K (corresponding
to $\approx$0.07\,$M_\odot$).
Therefore, for consistency across the full very-low-mass range
considered here, we adopted solar-metallicity (0 dex) atmospheres for
the main calculations.
However, we confirmed that the difference in magnitude along the
isochrone at the same mass between the 0 dex and $-$1 dex models is
small, with the $-$1 dex models yielding slightly larger magnitudes
but still remaining within $\lesssim$0.1 mag.

Reddening corrections are applied using the extinction law from
\citet{Wang2019}.
We first applied the extinction to each template spectrum and then
obtained bandpass-integrated extinctions using synphot by taking the
difference between the synthetic magnitude computed from the reddened
template and that from the corresponding unreddened template, after
integrating both through the full JWST bandpass. For the
synthetic-photometry validation, we interpolated within the BT
atmosphere grids in ($T_{\rm eff}$, $\log g$) to obtain representative
spectra prior to applying extinction and synphot.

The extinction values were evaluated at a representative mass of 0.1
$M_\odot$ along the 0.1 Myr isochrone. Although the reddening vectors
shown in the figures are computed at this mass, the bandpass
extinctions vary by $\le$0.01 mag across the full mass range of the
constructed isochrone, and the differences are therefore negligible.

In all color–magnitude and color--color diagrams presented in this
work, reddening vectors/arrows are derived using synthetic photometry
by applying extinction directly to the model spectra and recomputing
magnitudes through the full filter bandpasses.
Most sources have $A_V < 10$ mag \citep{Yasui2024}, based on
extinction estimates derived from NIRCam short-wavelength imaging of
the cluster region.
Accordingly, throughout this work we compute and annotate the
reddening arrows at the conservative upper bound $A_V=10$ mag, which
represents the upper envelope of the estimated extinction
distribution; hereafter, the ``extinction range considered'' refers to
$0 \le A_V \le 10$ mag. This choice is for annotation and consistency
only and does not affect our conclusions.

In particular, for the NIRCam long-wavelength and MIRI filters that
form the basis of our main analysis, we verified that the deviation
from a simple linear $A_\lambda /A_V$ approximation is negligible
($\le$0.01 mag) over the relevant extinction range. For the NIRCam
short-wavelength filters, where bandpass effects lead to a stronger
departure from linear extinction behavior, the reddening arrows should
be interpreted as representative, synthetic-photometry-based reference
displacements rather than as strictly linear extinction vectors.

Applied to Figure~\ref{fig:CM_MIRI_NIRCam}, from these extinction
estimates, at $A_V = 10$ mag the bandpass extinction and color-excess
values are $A_{\rm F356W} = 0.3$ mag and $E(m_{\rm F356W} - m_{\rm
  F770W}) = 0.03$ mag; we indicate these with red arrows in the
figures.
Because the impact of extinction is very small at these wavelengths,
the mass detection limit is set primarily by the F770W
sensitivity. From the 10$\sigma$ limiting magnitude in F770W and the
adopted isochrone tracks, the mass detection limit is estimated to be
$\simeq$0.03\,$M_\odot$. Disk excess emission can allow detection of
even lower-mass sources, particularly during early disk evolutionary
stages (Class I/II), as discussed in
Section~\ref{sec:dust_diskfraction}.

To supplement faint sources undetected in F770W, we also analyzed the
NIRCam long-wavelength (LW) color--magnitude diagrams,
$(m_{\mathrm{F356W}} - m_{\mathrm{F444W}})$ vs. $m_{\mathrm{F356W}}$
diagrams, shown in Figure~\ref{fig:CM_F356W_F444W}.  This diagram
provides {a means} of estimating mass detection limits for
low-mass objects for Cloud 2-N and Cloud 2-S. The dashed lines
indicate {the 10$\sigma$ detection limits} in the F356W and F444W
bands, the reddening vector for $A_V = 10$ mag {(upper bound of
  the extinction range)} is shown by red arrows, and {the
  isochrone model described above (0.1 Myr)} is plotted as blue
curves.
For $A_V = 10$ mag, the corresponding extinction values are
$A_{\mathrm{F356W}} \simeq 0.3$ mag and $E(m_{\mathrm{F356W}} -
m_{\mathrm{F444W}}) \simeq 0.09$ mag.
As in Figure~\ref{fig:CM_MIRI_NIRCam}, because the effect of
extinction is very small, the mass detection limit is set primarily by
the F356W sensitivity, and
the mass detection limit for the NIRCam bands is estimated to be
$\ll$0.017\,$M_\odot$. This limit is significantly deeper than that of
the F770W band, and allows for the detection of very low-mass stars
and brown dwarfs, which are discussed further in
Section~\ref{sec:disk_bd}.

For gas disk evolution (Section~\ref{sec:gas_disk}), we examined the
F405N band, which includes the Br$\alpha$ hydrogen emission line
($\lambda = 4.052$ $\mu$m).
Figure~\ref{fig:CM_F356W_F405N} shows the F405N color--magnitude
diagrams, $(m_{\mathrm{F356W}} - m_{\mathrm{F405N}})$
vs. $m_{\mathrm{F405N}}$ diagrams.
The dashed lines indicate the 10$\sigma$ detection limit in the F405N
band, and the isochrone model described above (0.1 Myr) is plotted as
blue curves.
The $A_V = 10$ mag reddening vector is indicated in red.
Based on these limits and the isochrone tracks, the mass detection
limit is estimated to be $\simeq$0.05\,$M_\odot$, equivalent to the
F770W band.

\subsection{MIR SED slope: Disk Classification} \label{sec:sed}
  For objects within the cluster regions defined by \citet{Yasui2024},
we derived the MIR SED slope, defined as $\alpha = d\log(\lambda
F_\lambda)/d\log(\lambda)$, using fluxes from the F356W, F444W, and
F770W bands. The slope was calculated for sources detected in the
F770W band.
In total, 89 objects in Cloud 2-N and 95 objects in Cloud 2-S were
detected in F770W and included in this analysis, as reported in
Section~\ref{sec:phot}.
Most of these sources were also detected in the F356W and F444W bands;
however, in rare cases where F444W data were unavailable, the slope
was derived using only F356W and F770W.

The SEDs for objects in the Cloud 2 cluster regions are shown in
Appendix~\ref{sec_appendix:SED} {(Figures~\ref{fig:SED_CL2N} and
  \ref{fig:SED_CL2S}), which provide full multi-band photometry from
  F115W to F770W along with the derived MIR SED slopes.}  In these
figures, data points from the F356W, F444W, and F770W bands are marked
with filled circles, and the derived slope is indicated by a black
line. For reference, NIRCam short-wavelength data (F115W, F150W,
F200W) with SNR $\geq$10$\sigma$ from \citet{Yasui2024} are also shown
as filled circles.

Based on the derived $\alpha$ values, disk evolutionary stages were
classified following the schemes of \citet{Lada2006} and
\citet{Hernandez2007}: $\alpha > 0$ for class I candidates, $-1.8 <
\alpha \le 0$ for class II stars, $-2.56 \lesssim \alpha <-1.8$ for
stars with evolved disks (EV), and $\alpha < -2.56$ for class III
(diskless) stars.

The boundary between evolved disks and Class III stars ($\alpha =
-2.56$) is adopted following \citet{Lada2006}. In their study, this
threshold was based on $\alpha = -2.66$ derived for M1-type stars
($\sim$0.4 $M_\odot$) by \citet{Hartmann2005}, with a margin of 0.1
applied to define the boundary.
\citet{Hernandez2007} pointed out that this boundary may vary slightly
for very late-type stars, based on trends observed in 8 $\mu$m
data. However, by comparing their results with the F770W magnitudes
used in this study, we find that the $\alpha = -2.56$ threshold
remains appropriate, even though our JWST observations include objects
with masses lower than M1-type stars. Therefore, we adopt $\alpha =
-2.56$ as a consistent and suitable boundary for disk classification
in the present analysis.

For the Cloud 2-N cluster, out of the {89} objects detected in the
F770W band, the numbers of Class I, II, EV, and III objects are
estimated to be 12, 50, 14, and 13, respectively. For the Cloud 2-S
cluster, out of 95 objects detected in F770W, the numbers are 23, 53,
13, and 6, respectively.
In Figures~\ref{fig:CM_MIRI_NIRCam}, \ref{fig:CM_F356W_F405N}, and
subsequent figures, the plots for Class I/II/EV/III objects are shown
by black filled circles enclosed in circles, black filled circles,
encircled plus symbols, and black open circles, respectively.

Figure~\ref{fig:CM_MIRI_NIRCam} presents the results of MIR SED-based
classification for sources detected in F770W.
The distribution of sources in the diagrams reveals clear separation
among disk classes, with Class I, Class II, EV, and Class III sources
occupying distinct regions.
As expected from the construction of the $\alpha$, which is based on
the same photometric bands, the $(m_{\rm F356W} - m_{\rm F770W})$
color index shows a strong correlation with $\alpha$.
The horizontal distribution in $(m_{\rm F356W} - m_{\rm F770W})$
corresponds to $\lesssim$0.3, $\simeq$0.3--1.0, $\simeq$1.0--2.5, and
$\gtrsim$2.5 mag for Class I, Class II, EV, and Class III sources,
respectively.
While this correlation is inherent to the method, the color-based
separation provides a visually intuitive representation of the
classification and helps to illustrate the distribution of disk types.

In Figure~\ref{fig:CC_MIRI_NIRCam}, we show the MIRI-NIRCam
color--color diagrams, $(m_{\mathrm{F444W}} - m_{\mathrm{F770W}})$
vs. $(m_{\mathrm{F356W}} - m_{\mathrm{F444W}})$, for the Cloud 2
clusters.
In this diagram, the main-sequence dwarf track is shown as a blue
curve.
The $A_V = 10$ mag reddening vector is indicated in red.
Note that, although extinction generally decreases with longer
wavelength, the reddening vector points slightly downward in
$(m_{\mathrm{F444W}} - m_{\mathrm{F770W}})$, reflecting the wavelength
dependence of extinction toward the 9.7 $\mu$m silicate feature (e.g.,
\citealt{Draine2003}), whose short-wavelength wing is sampled by the
broad F770W band.
By convention in disk-excess diagnostics and disk-fraction analyses
\citep[e.g.,][]{Lada1999, Haisch2000}, the photospheric reference in
color–color space is the unreddened dwarf locus; accordingly, we adopt
this track and consider masses down to 0.1 $M_\odot$.

Because standard main-sequence loci in the JWST photometric system are
not readily available, we adopted the 1 Gyr isochrone from
\citet{Baraffe2015} for masses $\ge$0.1 $M_\odot$. This choice is
motivated by the fact that evolutionary models indicate stars in this
mass range remain on the main sequence for much longer than 1 Gyr,
ensuring that the track represents stable main-sequence stars.
Instead of the D'Antona models \citep{{D'Antona1997}, {D'Antona1998}}
used to derive isochrone tracks in Section~\ref{sec:cm}, we use
\citet{Baraffe2015} because the D'Antona models do not extend to 1
Gyr.
To confirm consistency with the approach used to derive isochrone
tracks in Section~\ref{sec:cm}, we also computed synthetic magnitudes
in the JWST filter system by combining stellar parameters from
\citet{Baraffe2015} with BT-NextGen stellar atmosphere models
\citep{Allard2012}.
This verification showed that the resulting track closely matches the
track derived from the Section~\ref{sec:cm} approach, validating that
both methods yield nearly identical loci.
Throughout this paper, ``dwarf track'' denotes the locus of unreddened
main-sequence stars in the adopted JWST colors.

In this diagram, sources with infrared excess are clearly separated
from the dwarf track: the unreddened dwarf locus occupies the bluest
region and does not overlap the area populated by excess sources.
As in previous Spitzer/IRAC studies \citep[e.g.,][]{Lada2006}, disk
classes are clearly separated: Class III objects occupy the lower
left, evolved disks lie in the middle, and optically thick disks
(Class I and II) are found in the upper right.
This separation highlights the strong correlation between the NIRCam
and MIRI colors in these bandpasses, which is consistent with the disk
classification based on SED slopes adopted in this study.

Previous studies (e.g., \citealt{Gutermuth2009}) have noted that
extinction can affect disk classifications based on MIR SED slopes
($\alpha$) in regions of very high $A_V$, although the impact is
usually small. To verify this for our dataset, we calculated the
change in the $\alpha$ using the extinction law of \citet{Wang2019},
as described in Section~\ref{sec:cm}. Since most sources in the Cloud
2 clusters have $A_V \leq 10$ mag, this value serves as an upper limit
in our analysis. Under this condition, the difference between the
observed and extinction-corrected $\alpha$ is nearly constant at
$\Delta\alpha \approx -0.01$, regardless of the original $\alpha$
value. Therefore, extinction does not play a significant role in MIR
SED slopes, and we do not apply extinction correction for disk
classification.

This MIR SED slope--based classification, using F356W, F444W, and
F770W data---enables disk identification down to $\simeq$0.05
$M_\odot$. Notably, this study represents the first comprehensive
classification of circumstellar disks around such low-mass stars in a
low-metallicity star-forming cluster using MIR data up to 7.7
$\mu$m. By overcoming previous sensitivity limitations at wavelengths
$\geq$3$\mu$m, this approach allows for statistically robust analysis
of disk properties and their distribution across different
evolutionary stages.


\subsection{Identification of Cluster Members: $A_V$-limited sample}
\label{sec:ident_cluster}

To enable reliable statistical analyses of disk properties and
evolutionary stages, it is essential to first identify cluster member
candidates and minimize contamination from foreground
sources. \citet{Yasui2024} suggested that sources with low extinction
are primarily foreground sources. Following the same approach, we
plotted the sources with derived SEDs on the NIRCam short-band
color--magnitude diagrams, $(m_{\mathrm{F150W}} - m_{\mathrm{F200W}})$
vs. $m_{\mathrm{F200W}}$, in Figure~\ref{fig:cc_nircams}.  In this
diagram, {the 0.1 Myr isochrone track introduced in
  Section~\ref{sec:cm} is shown as a blue curve.}
The figures indicate that short NIRCam photometry in two bandpasses
alone cannot reliably distinguish between sources with intrinsic
infrared excess and those that are highly reddened sources;
longer-wavelength data, such as the F770W band used to derive the
$\alpha$ slope, break this degeneracy.

In this short-band color–magnitude diagram, most sources located at
small $m_{\rm F150W} - m_{\rm F200W}$---i.e., the locus corresponding
to low effective $A_V$ as determined by the short-wavelength NIRCam
data---are Class III (diskless) objects.
Although this could be attributed to the lack of disk excess in
Class III sources, the 2 $\mu$m disk excess is known to be very
small even for disk-bearing YSOs, so the short-wavelength color
primarily traces extinction rather than disk emission.
Consistent with the above, the short-band color--magnitude diagram
alone does not by itself provide a reliable disk classification, as
sources with a wide range of SED-based classes (Class I/II/EV/III)
overlap in this diagram. This mixing demonstrates that the small
$m_{\rm F150W} - m_{\rm F200W}$ color cannot be explained solely by
the absence of disk excess. By comparing with the longer-wavelength
classification (e.g., MIR slope $\alpha$), we find that the sources
clustered at small $m_{\rm F150W} - m_{\rm F200W}$ are predominantly
low-$A_V$ Class III objects, indicating that their small color is
primarily due to low extinction rather than the weakness of the 2
$\mu$m disk excess.

This clustering supports the previous suggestion by \citet{Yasui2024}
that low-$A_V$ sources can be considered foreground objects unrelated
to ongoing star formation activity.  Here, we define the boundary by
shifting
the isochrone track along the extinction vector, as shown by dotted
blue curves in Figure~\ref{fig:cc_nircams}, so that its shifted
position defines a boundary separating most Class III sources from the
cluster member candidates.
Following the synphot-based bandpass-extinction procedure described in
Section~\ref{sec:cm}, we shift the isochrone along the reddening
trajectory to construct this boundary.
The amount of shift corresponds to an extinction of approximately
$A_V=2.5$ and 1 mag in the Cloud 2-N and 2-S cluster regions,
respectively.
Sources located to the right of this boundary are considered cluster
member candidates, as disk-bearing sources are predominantly
concentrated in this region.
Sources located to the left of this boundary are regarded as
foreground objects and are shown with gray symbols in the figure,
while sources to the right of the boundary are defined as the
``$A_V$-limited sample'' in this study.

In \citet{Yasui2024}, the $A_V$-limited sample was defined based on
the extinction-shifted isochrone tracks from the D'Antona model
\citep{{D'Antona1997}, {D'Antona1998}}, with the boundary determined
from the $A_V$ distribution in the cluster region and the control
field. This boundary was set under the assumption that sources located
to the right of the extinction-shifted isochrone are likely embedded
in the molecular cloud and thus considered cluster member
candidates. In contrast, our boundary is defined using a different
approach—based on the distribution of Class III sources and the
concentration of disk-bearing objects.

Despite these independent methods, we find that the sources
identified as the $A_V$-limited sample in \citet{Yasui2024} and
those selected by our method show strong agreement. In the Cloud 2-N
cluster region, six sources classified as cluster members in
\citet{Yasui2024} are excluded in our sample, while in the southern
region the member candidate identifications are fully consistent
between the two studies.

This difference is likely due to slight variations in the shape of
the 0.1 Myr isochrone track used in each study. In addition, the
adopted $A_V$ boundary differs slightly: \cite{Yasui2024} used $A_V
= 2$\,mag for the Cloud 2-N cluster, whereas we adopted
$A_V=2.5$\,mag, while the Cloud 2-S cluster uses the same $A_V =
1$\,mag in both studies.  However, these small differences do not
significantly affect the overall consistency between the two
selection methods.
A direct comparison between our Figure~\ref{fig:cc_nircams} and
Figure~9 in \citet{Yasui2024} highlights the consistency between the
two selection approaches and reinforces the robustness of the
$A_V$-limited sample definition.

The number of sources included in the $A_V$-limited sample selected by
this method is 3/45/13/2 and 5/49/9/2 for Class I/II/EV/III in the
Cloud 2-N and 2-S cluster regions, respectively.

\subsection{Mass-$A_V$-limited sample and Disk Fraction Estimation}
\label{sec:dust_diskfraction}

To ensure a fair and unbiased comparison across disk classes, we adopt
a consistent mass threshold based on the detection limit of Class III
objects. Unlike Class I and Class II sources, Class III objects lack
significant infrared excess and thus provide a more conservative and
reliable baseline for mass-limited detection. As discussed in
Section~\ref{sec:cm}, while the nominal 10$\sigma$ detection limit in
the F770W band corresponds to $\simeq$0.03\,$M_\odot$ on the 0.1 Myr
isochrone, we define the red line in Figure~\ref{fig:cc_nircams} at
the position corresponding to $\simeq$0.04\,$M_\odot$, which roughly
matches the faintest detected Class III or EV sources and thus serves
as a practical threshold for the mass-limited sample.

Figure~\ref{fig:cc_nircams} also shows that, for sources other than
Class III, objects with masses below this nominal detection limit are
detected. This is because Class I and Class II sources exhibit strong
infrared excess in the F770W band, which enhances their detectability
even at lower masses. As shown in Figure~\ref{fig:CM_MIRI_NIRCam},
Class I and Class II sources typically exhibit excesses of
$\gtrsim$2.5 mag and $\gtrsim$1.0 mag in $m_{\mathrm{F356W}} -
m_{\mathrm{F770W}}$, respectively. Indeed, some of these sources are
detected even when they are up to $\sim$2 mag fainter than the red
boundary.
If the disk fraction were calculated using all sources for which
$\alpha$ values are derived, the result would be biased toward these
early-stage objects, as their enhanced infrared emission allows them
to be detected even below the nominal mass threshold.
To avoid this bias and ensure a fair comparison across disk classes,
we apply the Class III-based mass threshold uniformly to all disk
classes and define sources located above the red mass-selection
boundary in Figure~\ref{fig:cc_nircams} as the ``Mass-limited
sample.''

The red boundary is anchored at the 0.04 $M_\odot$ position on the 0.1
Myr isochrone and extended along the reddening direction by applying
bandpass-integrated extinctions derived from synthetic photometry in
the relevant NIRCam filters (F150W, F200W)
across the extinction range considered in this study; interpolation
between the computed points yields a continuous boundary. Because the
vast majority of sources have $A_V \le 10$ mag, we adopt $A_V = 10$
mag as the representative extinction range when defining this
reddening displacement. A small number of more highly extinguished
objects ($A_V \simeq 20$--30 mag) appear in the color--magnitude
diagram, and the boundary is extended to this upper range for
completeness.

However, we verified that computing the reddening
displacement using $A_V=10$, 20, and 30 mag individually yields an
essentially identical trajectory to that obtained by extending the
$A_V = 10$ mag vector, because the reddening direction is preserved
across this extinction range and only the displacement length
changes with $A_V$.
This behavior reflects a general property of the NIRCam
short-wavelength filters: despite modest non-linearities in
$A_\lambda/A_V$ at higher extinctions, the reddening direction remains
essentially unchanged. Consequently, the same conclusion applies to
all NIRCam SW color–magnitude diagrams presented later in this work.
As a result, the source selection remains unchanged, and this has no
impact on the disk-fraction analysis presented later in this work.

When combined with the $A_V$-limited sample defined in the previous
section, the resulting subset is referred to as the
``Mass-$A_V$-limited sample.'' This sample is designed to minimize
both extinction and mass biases, thereby enabling a more robust
statistical analysis of disk populations.

The number of sources satisfying these conditions in the Cloud 2-N
cluster is 2, 40, 12, and 2 for Class I, Class II, EV, and Class III,
respectively. In the Cloud 2-S cluster, the corresponding numbers are
5, 43, 8, and 2.
Following the method in \citet{Megeath2012}, the disk
fraction---defined as the number of Class II sources divided by the
total number of Class II, EV, and Class III sources---{was previously
  estimated using the method employed in \citet{Haisch2000}, where
  uncertainties were derived from Poisson statistics applied only to
  the disk-bearing sources. In this study, we adopt a more rigorous
  error propagation method that accounts for uncertainties in both
  counts, providing a statistically more accurate
  estimate\footnote{The uncertainty is calculated using the
  propagation formula: $\sigma_p = p \times \sqrt{\frac{1}{a} +
    \frac{1}{b}}$, where $p = a/b$, $a$ is the number of disk-bearing
  sources, and $b$ is the total number of sources considered.}.
Using this approach, the disk fraction is 74$\pm$15\% (40/54) for
Cloud 2-N and 81$\pm$17\% (43/53) for Cloud 2-S.

Although this Mass-$A_V$-limited sample excludes sources not detected
in the F150W and F200W bands from the disk fraction calculation, this
exclusion is expected to have minimal impact.
Nearly all sources, except Class I objects, are detected in at least
one of these bands, and most are detected in both. In the Cloud 2-N
cluster, only six sources (five Class II and one Class III) are not
detected in the F150W and/or F200W bands. In the Cloud 2-S cluster,
only five sources (four Class II and one EV source) are not detected
in these bands. Therefore, the disk fraction values presented here
should be interpreted as lower limits.

To assess the completeness of F770W detections, we also plot sources
detected in the NIRCam bands but not in the F770W band as small dots
in Figure~\ref{fig:cc_nircams}.
In the Cloud 2-N cluster, only three sources are detected in the
NIRCam bands but remain undetected in the F770W band, whereas in Cloud
2-S, as many as 13 such sources are present.
Their non-detection in the F770W band---especially in Cloud 2-S, where
the stellar density is higher---does not appear to be driven solely by
intrinsic faintness, and is instead consistent with being affected by
observational limitations such as source confusion.
The completeness of F770W detections for the Mass-$A_V$-limited sample
is estimated as 94\% (54/(54+3)) for Cloud 2-N and 80\% (53/(53+13))
for Cloud 2-S.

\section{Discussion} \label{sec:discussion}

\subsection{Dust Disk Evolution} \label{sec:dust_disk}
\subsubsection{Dust Disk Evolution for Cloud 2 Clusters}
\label{sec:dust_disk_cl2}

In the preceding sections, we analyzed the disk categorization within
Cloud 2 clusters using observations up to 7.7 $\mu$m. Observations at
7.7 $\mu$m allow for the identification of both optically thick and
optically thin disks \citep{Sicilia-Aguilar2006}, contributing to a
more complete view of disk-bearing sources. To further investigate
disk evolution, we now examine shorter wavelengths that trace the
inner and hotter regions of circumstellar disks.  Observations in the
2--4 $\mu$m range {primarily} trace the hottest inner disk regions
and are sensitive only to optically thick material. By comparing
results across these wavelength ranges, we aim to achieve a more
comprehensive understanding of disk evolution in the Cloud 2 clusters.

First, we compare the disk classification obtained in this study with
the determination of disk presence based on NIR data up to
2 $\mu$m. For this comparison, we use JHK-band data obtained with the
MKO filter system from previous observations \citep{Yasui2009}. The
MKO system provides a well-established framework for identifying
disk-bearing sources based on NIR colors
and for deriving disk fractions using color–color diagrams (see
details in Appendix~\ref{sec_appendix:NIRobs}, which describes the
comparison procedure, the adopted MKO criteria, and the coordinate
matching method).

Importantly, the detection limit in terms of stellar mass and the
angular resolution of the MKO JHK data (Subaru/MOIRCS) are broadly
comparable to those of our MIRI/F770W imaging, both reaching
approximately 0.1 $M_\odot$ and ${\rm FWHM} \simeq 0\farcs3$; thus,
the NIR–MIR comparison is not driven by substantial differences in
depth or spatial resolution.

We emphasize that JWST/NIRCam F200W is not equivalent to the classical
K/Ks band.  The F200W filter has an effective wavelength of
$\simeq$1.99 $\mu$m, which is significantly shorter than that of the
classical K/Ks bands ($\simeq$2.15--2.2 $\mu$m).
As disk excess emission generally increases with wavelength, the
already weak excess at $\simeq$2 $\mu$m becomes even smaller when
measured in F200W, where the stellar photosphere contributes more
strongly relative to the disk emission. In addition, the empirical
dwarf locus and reddening vector in the JWST NIR filters are still
being consolidated; for these reasons, a robust and directly
comparable definition of a 2 $\mu$m excess boundary in the JWST/NIRCam
F200W system is not attempted in this work.

As discussed in Section~\ref{sec:ident_cluster}, disk excess emission
at 2 $\mu$m is intrinsically small, particularly for low-mass stars
and brown dwarfs. As a result, the inferred disk fraction depends
strongly on the exact placement of the dwarf locus and the reddening
vector in color--color space; even a modest shift in the adopted
decision boundary can lead to a substantial change in the derived disk
fraction. This strong dependence makes the choice of photometric
system and empirical calibration especially critical at 2 $\mu$m.

Accordingly, we retain the MKO-based determination for this specific 2
$\mu$m comparison, while using JWST primarily for disk diagnostics at
longer wavelengths ($\ge$3 $\mu$m), where disk excess emission is
substantially larger and far less sensitive to small uncertainties in
the adopted dwarf locus or reddening vector \citep{Haisch2001AJ}.
Although JWST offers superior sensitivity and angular resolution,
standardized and empirically validated criteria specifically for
identifying disk excess at 2 $\mu$m in the JWST/NIRCam filter system
are still being established. In contrast, the MKO JHK system has been
widely used in previous studies, with well-tested and empirically
anchored methods for defining near-infrared disk excess and estimating
disk fractions. For this reason, and to ensure consistency and
robustness in the 2 $\mu$m comparison, we adopt the MKO‑based
determination in this work.

The matched sources are plotted on the NIR JHK color--color diagrams
(Figure~\ref{fig:cc_mcs}), with different symbols according to the
disk classification based on $\alpha$ from this study. In
Figure~\ref{fig:cc_mcs}, the lower right side of the gray dot-dashed
line is considered the disk excess region, while the upper left side
is considered the disk-free region.
As a result, all Class III objects are found to be distributed in the
disk-free region, while some Class I objects are located in the disk
excess region. Specifically, in Cloud 2-N, none of the 3 Class I
objects fall in the disk excess region, while in Cloud 2-S, 4 out of 7
Class I objects do. For Class II objects, 4 out of 29 in Cloud 2-N and
9 out of 36 in Cloud 2-S are located in the disk excess region, with
the majority in both clouds found in the disk-free region, suggesting
a lack of detectable inner disk excess at NIR wavelengths.
Additionally, most of the objects classified as Class I in this study
are not plotted in this diagram because they are not detected at 1--2
$\mu$m.

As shown in Appendix~\ref{sec_appendix:SED}, most Class I sources are
not detected below 2 $\mu$m, particularly at $\le$1.5 $\mu$m (Cloud
2-N: 9/12; Cloud 2-S: 18/23). These non‑detections further limit the
usefulness of a strictly 2$\mu$m metric in JWST/NIRCam for embedded
sources. Accordingly, we retain the MKO JHK-based fraction for the 2
$\mu$m comparison, while using JWST---NIRCam at 3, 4 $\mu$m and MIRI
at 7 $\mu$m---to classify disks and to derive disk fractions at 3, 4,
and 7 $\mu$m in this paper.

Next, we examine disk excess at longer NIR wavelengths using data up
to F356W and F444W. Disk excess at 3 $\mu$m is diagnosed using the
$(m_{\mathrm{F200W}} - m_{\mathrm{F356W}})$ vs. $(m_{\mathrm{F115W}} -
m_{\mathrm{F150W}})$ diagram (Figure~\ref{fig:CC_NIRCam_F356W}),
{while disk excess at 4 $\mu$m is diagnosed using the
  $(m_{\mathrm{F200W}} - m_{\mathrm{F444W}})$ vs. $(m_{\mathrm{F115W}}
  - m_{\mathrm{F150W}})$ diagram (Figure~\ref{fig:CC_NIRCam_F444W}).}
Based on previous conventions \citep[e.g.,][]{Lada1999, Haisch2000}, a
boundary line parallel to the reddening vector passing through the
position of the main-sequence dwarf track corresponding to a late-type
M star (approximately 0.1 $M_\odot$), introduced in
Section~\ref{sec:cm}, is used.
This line, shown as a gray dot-dashed curve in the figure, separates
the disk-excess region (upper left) from the disk-free region (lower
right).

In contrast to the 2 $\mu$m case, the separation between disk-bearing
and disk-free sources becomes much clearer at these longer
near-infrared wavelengths. In both Figures~\ref{fig:CC_NIRCam_F356W}
and \ref{fig:CC_NIRCam_F444W}, sources classified as hosting disks
based on their F770W emission occupy regions of the diagrams that are
well separated from disk-free sources, illustrating that disk excess
at $\ge$3 $\mu$m is both stronger and less sensitive to the exact
placement of the decision boundary.

In both figures, the objects are plotted using the same symbols as in
Figure~\ref{fig:CM_MIRI_NIRCam}, based on the disk classification
determined by the MIR SED slope $\alpha$ in Section~\ref{sec:sed}.
For F356W, the disk fraction is 68$\pm$14\% (38/56), for Cloud 2-N and
73$\pm$14\% (46/64) for Cloud 2-S.
For F444W, the disk fraction is 79$\pm$16\% (44/56) for Cloud 2-N and
79$\pm$15\% (50/63) for Cloud 2-S.
Here, the disk fraction is defined as the number of Class II sources
divided by the total number of Class II, evolved disk, and Class III
sources, following the same definition used for the 7 and 2 $\mu$m
case described earlier.
These figures illustrate that the boundary generally separates objects
with optically thick disks (Class I + Class II) from those with no
disks or optically thin disks (Class III + evolved disk sources),
although systematic deviations exist: in the F356W diagram, some Class
II sources fall in the disk-free region, leading to an underestimate
of the disk fraction, whereas in the F444W diagram, most Class II
sources lie in the disk excess region but some evolved disk sources
also fall there, resulting in a tendency to overestimate the disk
fraction.

In summary, for the Cloud 2 clusters, the disk fraction traced at
$\geq$3 $\mu$m is approximately consistent with the optically thick
dust disk fraction determined from the MIR SED slope ($\alpha$)
($\sim$70--80 \%), while the disk fraction traced at 2 $\mu$m is
significantly lower ($\sim$10--30 \%). This finding implies that, even
at such an early evolutionary stage, the innermost disk excess is
already diminished or undetectable in a large fraction of sources.

\subsubsection{Comparison with Previous Studies for nearby regions}

This section presents a comparative analysis of dust disk evolution in
the Cloud 2 clusters, as discussed in Section~\ref{sec:dust_disk_cl2},
with results from previous studies of nearby star-forming regions. By
examining disk fractions derived from observations at various
wavelengths, we aim to assess whether similar evolutionary trends are
seen in low-metallicity environments compared to solar-metallicity
clusters.

To begin this comparison, we first review disk evolution trends
observed in nearby solar-metallicity clusters based on MIR data,
particularly those obtained with Spitzer. Studies of disk evolution
using SED slopes derived from observations up to 8 $\mu$m have been
primarily conducted with Spitzer in the solar neighborhood. As a
result, it has generally been reported that younger regions show
higher disk fractions, which decline over approximately 10 Myr with
increasing age \citep[e.g.,][]{Hernandez2007}. In young regions such
as NGC 2024 and the Orion Nebula Cluster (ONC), with ages $\lesssim$1
Myr, the disk fraction is estimated to be around $\sim$75\%
\citep{Megeath2012, Megeath2016}.

The high disk fraction observed in the Cloud 2 clusters, based on
observations up to 7.7 $\mu$m, is consistent with these Spitzer-based
results for very young solar-metallicity clusters.  This agreement
suggests that similar disk evolution occurs even in the
low-metallicity star-forming clusters targeted here.

In contrast, when using shorter wavelength data (up to 3--4 $\mu$m),
previous studies of nearby star-forming regions (e.g.,
\citealt{Lada2006}; \citealt{Hernandez2007}) have noted that disk
detection may be less reliable than at longer wavelengths.
This reduced reliability may lead to an underestimation of disk
fractions, or may simply reflect the intrinsically lower disk excess
at these wavelengths.
Some studies (e.g., \citealt{Ribas2014}) have also suggested that
shorter disk lifetimes could contribute to the observed low disk
fractions, although the extent of this effect remains uncertain.
Nevertheless, disk classification based on data up to 3--4 $\mu$m is
found to be approximately consistent with that based on $\alpha$
derived from 7.7 $\mu$m observations (e.g., \citealt{Ribas2014};
\citealt{Gutermuth2009}; \citealt{Megeath2012}.
This consistency is also observed in the Cloud 2 clusters.

However, the situation differs when considering data at 2
$\mu$m. Although NIR observations at this wavelength—corresponding to
the K-band—are generally insufficient to reliably determine the
presence or absence of circumstellar disks, they can still provide
useful insights into overall trends within a cluster. For example,
\citet{Haisch2000} conducted imaging observations up to 3 $\mu$m
(L-band) for NGC 2024, a very young star-forming region (age
$\simeq$0.3 Myr) similar to Cloud 2, and pointed out that K-band data
alone do not allow for robust disk identification. Nevertheless, they
found that approximately 58 \% of the total sources—corresponding to
about 70 \% of those identified as disk-bearing in the L-band—were
located in the disk excess region on the JHK color--color diagram.

Similarly, \citet{Yasui2009} analyzed K-band imaging data of NGC 2024
obtained with Subaru/MOIRCS, using the same instrument and reduction
procedures as for the Cloud 2 clusters. The resulting NIR disk
fraction for NGC 2024 was found to be high, at 65\%, consistent with
previous studies (see Appendix in \citealt{Yasui2009}).
In addition, based on results from nearby star-forming regions,
\citet{Yasui2009} compared disk fractions derived from JHK data alone
with those from JHKL data. They found that the JHK-only disk fractions
are approximately 60\% of the JHKL-based values, with slightly more
scatter in the disk fraction–age relation. Nevertheless, the estimated
disk lifetime derived from JHK data is essentially identical to that
from JHKL data. These findings indicate that, despite somewhat larger
uncertainties, JHK-only data can still capture the overall
evolutionary trend of decreasing disk fraction with age.

In contrast, the Cloud 2 clusters show a markedly different trend:
despite their very young age, significantly lower disk fractions are
derived from 2 $\mu$m data ($\sim$10--30 \%), even at the earliest
evolutionary stages.  This suggests that, in these low-metallicity
environments, the excess emission from the innermost disk regions may
already be absent or substantially reduced, making 2 $\mu$m data less
effective for identifying disk-bearing sources, while still useful for
examining overall disk-related trends.

The detection rate of Class I objects at 2 $\mu$m also shows a
contrasting trend between Cloud 2 and nearby regions.
In nearby star-forming regions such as NGC 2024 and Taurus, most Class
I objects are detectable at 2 $\mu$m \citep[e.g.,][]{Haisch2000,
  Hartmann2005}, whereas in the Cloud 2 clusters, many Class I sources
are not detected at these wavelengths.

In summary, the disk fractions inferred from diagnostics at
wavelengths $\ge$3 $\mu$m in the Cloud 2 clusters are consistent with
previous studies in the solar neighborhood, {a large fraction of
  objects ($\sim$70--80\%)} show sufficient disk excess to be
classified as having optically thick disks. However, at 2 $\mu$m,
the fraction is systematically lower---only $\sim$10--30\% of sources
show enough excess to meet the optically thick criterion, even among
objects classified as optically thick at 7.7 $\mu$m---indicating a
wavelength-dependent sensitivity to inner-disk emission.
This discrepancy suggests that the innermost disk material traced at 2
$\mu$m may have already dissipated or become undetectable, even in
objects that show clear disk excess at longer wavelengths.
Various factors may explain why disks traced only at 2 $\mu$m are not
observed, such as disk dissipation, dust growth or settling, dust
sublimation, and others, but this is discussed in
Section~\ref{sec:diskevolve_lowmeta}.

\subsection{Implication to Gas Disk Evolution} \label{sec:gas_disk}

We investigate gas disk evolution using the {\rm NIRCam} narrow-band
filter F405N, which covers the hydrogen recombination line Br$\alpha$
($\lambda = 4.05$ $\mu$m). Accreting gas is originally traced by UV
and the H$\alpha$ line in the optical, and the classification of
CTTS/WTTS is based on the equivalent width (EW) of the H$\alpha$
emission line (10 \AA~criterion; \citealt{Appenzeller1989}). While
mass accretion rates are often derived from H$\alpha$ spectroscopy,
narrow-band imaging has also been used for this purpose (e.g.,
\citealt{{Drew2005}, {Barentsen2011}}).
In $R–I$ vs. $R–{\rm H}\alpha$ diagrams, constructed using broadband R
and I filters (where the R band includes the H$\alpha$ line and the I
band covers a slightly longer wavelength range beyond H$\alpha$),
sources that greatly exceed the isochrone track in the $R-{\rm
  H}\alpha$ color are considered to be objects with mass accretion. In
addition, the mass accretion rate can be estimated from the amount of
the corresponding H$\alpha$ EW on the color--color diagram.

Following a similar approach, we constructed {Br$\alpha$-based}
color--color diagrams using F405N together with F356W (continuum
shortward of Br$\alpha$) and F444W (which {covers} Br$\alpha$):
the $(m_{\rm F356W} – m_{\rm F444W})$ vs. $(m_{\rm F356W} – m_{\rm
  F405N})$ diagrams (Figure~\ref{fig:CC_f405n_f356w_f444w}), hereafter
referred to as the Br$\alpha$ color--color diagram.
Dwarf tracks for the mass range of 0.1--1.4 $M_\odot$ are shown as
blue curves.
The $A_V$-limited sample is plotted, with the symbols the same as in
Figure~\ref{fig:CM_MIRI_NIRCam}, but Class I sources without NIRCam
short-wavelength (SW) detections (and thus without $A_V$ estimates)
are also included, as they are likely cluster members.

In these diagrams, dust disk excess shifts sources toward the upper
right, while Br$\alpha$ emission from gas accretion shifts them upward
relative to the intrinsic stellar color. To quantify these effects, we
performed synthetic photometry using the NIRCam transmission curves
and computed bandpass-convolved Vega magnitudes. We adopted a
BT-NextGen model for a 0.1 $M_\odot$ main-sequence star as the
fiducial photosphere and added a simple inner-disk continuum
parameterized by a power-law slope $\alpha_{\rm disk}$ and a
normalization $k \equiv F_{\rm disk}(1.5\,\mu{\rm
  m})/F_\star(1.5\,\mu{\rm m})$. For a given target MIR slope,
$\alpha_{\rm total}$ (monochromatic three-point at 3.565/4.402/7.639
$\mu$m), we solved for $k$ so that the composite SED $F_{\rm
  tot}=F_\star+k\,F_{\rm disk}$ attains $\alpha_{\rm total}$, and then
evaluated the colors $(m_{\rm F356W}-m_{\rm F444W},\, m_{\rm
  F356W}-m_{\rm F405N})$.

For physical consistency across the bulk of the sample, we fixed
$\alpha_{\rm disk}=1.2$ for the markers at $\alpha_{\rm
  total}=-2.56,-1.8,0$ and varied only $k$. To span the small number
of observed sources in the high-$\alpha$ tail, we extended the markers
to $\alpha_{\rm total}=2$ and $3$ using $\alpha_{\rm disk}=2.5$ and
$3.5$, respectively; these are illustrative extremes rather than
representative of the population.  The qualitative displacement
pattern is insensitive to stellar mass: repeating the calculations
over $0.1$--$1.4\,M_\odot$ (the range spanned by the dwarf track)
yields similar vectors, so the $0.1\,M_\odot$ case is representative
of low-mass members. We also verified that varying $\alpha_{\rm disk}$
within $[0.5,\,2.0]$ produces envelopes enclosing the $\alpha_{\rm
  total}\le 0$ markers (not shown), indicating that our conclusions
are robust to the assumed disk slope in the typical regime.
The synthetic-photometry locus for $-3\lesssim \alpha_{\rm total}\le
3$ is shown as the dark gray curve in
Figure~\ref{fig:CC_f405n_f356w_f444w}, anchored at the $0.1\,M_\odot$
dwarf track.

In Figures~\ref{fig:CC_f405n_f356w_f444w}, the dwarf track at 0.1
$M_\odot$ corresponds to an intrinsic photospheric SED slope of
$\alpha = -3.1$, as derived from the synthetic stellar spectrum, and
we use this point as the anchor for defining class boundaries. Orange
markers indicate the adopted MIR SED slope boundaries: Class III---EV
($\alpha = -2.56$), EV---Class II ($\alpha = -1.8$), and Class
II---Class I ($\alpha = 0$).

The effect of interstellar reddening is illustrated by the red arrow,
which was derived by applying extinction ($A_V = 10$ mag) directly to
the synthetic spectra prior to filter convolution. In the Br$\alpha$
color--color diagram, reddening produces a displacement of 0.06 mag in
the vertical direction and 0.09 mag in the horizontal direction.
The $\alpha$-induced displacement, shifted according to this reddening
vector, is shown as the dashed gray curve in
Figures~\ref{fig:CC_f405n_f356w_f444w}.

To quantify the effect of Br$\alpha$ line emission itself, we added a
Gaussian Br$\alpha$ emission line with a FWHM of 200\,km\,s$^{-1}$---a
typical width observed in hydrogen recombination lines of low-mass
young stellar objects (e.g., \citealt{Folha2001})---to the fiducial
stellar spectrum and recomputed the synthetic photometry. For an
equivalent width of 10 \AA(the classical CTTS/WTTS threshold), the
magnitude change is small—about 0.02 mag in F405N and $<$0.01 mag in
F356W and F444W—comparable to typical photometric uncertainties
($\sim$0.03 mag), making such weak Br$\alpha$ emission difficult to
detect reliably in imaging alone.

We therefore focus on a stronger emission case with an equivalent
width of 50 \AA. In this case, the synthetic calculations yield a
brightening of 0.12 mag in the F405N band, whereas the magnitude
changes in the F356W and F444W bands remain $<$0.01 mag.
This behavior is essentially identical for all stellar masses along
the dwarf track (0.1--1.4 $M_\odot$).
Consequently, Br$\alpha$ emission produces an almost purely vertical
displacement in the Br$\alpha$ color–color diagram. A light-gray
boundary line is defined by shifting the reddening-adjusted
$\alpha$-displacement vector in the direction corresponding to ${\rm
  EW} = -50$ \AA; sources above this boundary are classified as
exhibiting Br$\alpha$ excess.

To evaluate how this criterion relates to the actual mass accretion
rate, a quantitative relationship between Br$\alpha$ line strength and
accretion rate is required. Although Br$\alpha$ lies at the
long-wavelength end of the L-band and can be observed from the ground,
such observations are limited, and few YSO spectra provide both Br$\alpha$
measurements and known accretion rates.
\citet{Evans1987} reported that the Br$\alpha$ line has approximately
the same flux as the Br$\gamma$ line, suggesting that existing
Br$\gamma$–accretion rate relations may be used as a proxy for
estimating Br$\alpha$-based accretion rates.
According to \citet{Muzerolle1998}, although an equivalent width of
${\rm EW}= -50$ \AA{} lies outside the plotted range of the Br$\gamma$
EW–mass accretion rate relation, it can be extrapolated to correspond
to a mass accretion rate of $\gtrsim$10$^{-6}$ $M_\odot$\,yr$^{-1}$.
The fact that nearly all detected sources are low-mass stars, and that
approximately one-third of the Class I/II sources show Br$\alpha$
excess corresponding to ${\rm EW} \simeq -50$ \AA{}, suggests that
their mass accretion rates are $\gtrsim$10$^{-6}$
$M_\odot$ yr$^{-1}$.
This is a relatively large value, indicating that these low-mass stars
are still in an early phase of significant mass buildup.

Using these boundaries, it was found that a significant fraction of
Class I sources, particularly in the Cloud 2-S cluster, and a smaller
fraction of Class II sources lie above the light gray boundary lines.
The fractions of objects located above the boundary—where a gas disk
is considered to be present—are estimated to be 20$\pm$6 \% (12/61)
in the Cloud 2-N cluster and 28$\pm$7 \% (18/65) in the Cloud 2-S
cluster.
Furthermore, sources with larger $\alpha$ values, corresponding to
redder $(m_{\rm F356W} – m_{\rm F444W)}$ colors, are more likely to
retain their gas disks, as they tend to lie above the light gray
boundary in the Br$\alpha$ color–color diagrams.

The result that almost all of the detected sources are low-mass stars,
and that approximately one-third of these Class I/II sources are
thought to have gas disks indicating mass accretion rate of
$\gtrsim$10$^{-6}$ $M_\odot$ yr$^{-1}$, suggests that their mass
accretion rates are comparable to or slightly higher than those of
low-mass stars of similar age to the Cloud 2 clusters ($\sim$0.1 Myr)
in the solar neighborhood (e.g., top panel of Figure~8 in
\citealt{Hartmann2016}).

\subsubsection{Relationship Between Dust and Gas Disks in the Cloud 2 Clusters}

To further understand the nature of protoplanetary disks in the Cloud
2 clusters, we examine the relationship between sources showing
NIR/MIR (2--7.7 $\mu$m) excess and those exhibiting Br$\alpha$ excess
identified in Section~\ref{sec:gas_disk}.
Section~\ref{sec:dust_disk} focused on dust disk evolution and
revealed that a significant fraction of sources exhibit excess
emission at 3, 4, and 7.7 $\mu$m, comparable to those in nearby
solar-metallicity regions of similar age. However, at 2 $\mu$m, most
of the sources with 3--7.7 $\mu$m excess do not show corresponding
excess, indicating a distinct trend from nearby regions.

We first investigate how Br$\alpha$ excess sources are distributed
across evolutionary classes defined using MIR SED slope ($\alpha$).
In Cloud 2-N, the fractions are Class I: 20\% (1/5), Class II: 26\%
(11/43), EV: 0\% (0/10), and Class III: 0\% (0/3),
while in Cloud 2-S they are Class I: 71\% (10/14), Class II: 17\%
(7/42), EV: 14\% (1/7), and Class III: 0\% (0/2).
These ratios indicate that Br$\alpha$ excess---interpreted as a tracer
of active mass accretion---is most frequent among Class I objects,
less common among Class II and EV sources, and absent in Class III.

We further investigate how the presence of Br$\alpha$ excess relates
to the 2 $\mu$m excess. In Figure~\ref{fig:cc_mcs}, sources with
Br$\alpha$ excess are highlighted with cyan squares.
Among the Br$\alpha$ excess sources, some are located in the
lower-right region of the gray dotted borderline, indicating the
presence of 2 $\mu$m excess, while others are found in the upper-left
region, suggesting its absence.

In summary, Br$\alpha$ excess sources partially overlap with the
population showing MIR excess, but their correspondence with 2 $\mu$m
excess is not consistent. This suggests that Br$\alpha$ emission,
which traces gas disk components, does not always coincide with NIR
excess at shorter wavelengths.
{Importantly, the Br$\alpha$ detections in this study rely on a
  relatively high threshold (${\rm EW} = -50$\,\AA), which is
  substantially larger than the commonly adopted 10\,\AA{}
  criterion. As a result, many accreting sources with weaker
  Br$\alpha$ emission are likely to remain undetected. Furthermore,
  the color--color diagrams do not include information on the absolute
  brightness of the sources and therefore do not reflect the
  sensitivity limits relevant for Br$\alpha$ detection. Taken
  together, these observational limitations may contribute to the
  imperfect correspondence between Br$\alpha$ excess and 2 $\mu$
  excess.}

\subsection{Other Properties of Disk-Classified Sources within the Cloud 2 Clusters} \label{sec:other_prop}

\subsubsection{SED Slope Distributions and Evolutionary Implications}
\label{sec:sed_dist}

The distribution of $\alpha$ values is shown in
Figure~\ref{fig:sed_dist}.
Figure~\ref{fig:sed_dist} shows the distributions for the full,
$A_V$-limited, and Mass-$A_V$-limited samples as white, light gray,
and dark gray histograms, respectively.
In previous sections, the discussion has primarily focused on the
$A_V$-limited and Mass-$A_V$-limited samples. However, in this
section, which examines the distribution of SED slopes, we include
Class I sources that are not part of these samples.
Extinction estimates are derived using short-wavelength NIR data,
specifically the F150W and F200W filters of the NIRCam SW
channel. These wavelengths enabled the detection of Class III and EV
sources. Based on their estimated $A_V$ values, these sources were
excluded from the $A_V$-limited sample, as their low extinction
suggests they are likely foreground objects rather than cluster
members.
In contrast, Class I sources are not detected at short NIR wavelengths
and are therefore {also} not included in the $A_V$-limited
sample. Nevertheless,
Class I sources located within the cluster regions are considered
highly probable cluster members. Given the rarity of Class I objects
and their association with the earliest stages of star formation, it
is extremely unlikely that such sources would appear coincidentally
within the cluster boundaries. Their spatial concentration strongly
suggests a physical association with the cluster.

This figure shows that the SED slope distributions for both clusters
exhibit distinct peaks at approximately $\alpha \simeq -$2.8,
$\simeq$$-$1.0, and $\simeq$1.0, indicating that the sources can be
broadly categorized into three evolutionary groups.
Similar distributions have been reported in nearby star-forming
regions, where a clear separation between sources with $\alpha \simeq
-$2.8 and $\sim$$-$1.0 has been interpreted as a transition between
diskless stars (Class III) and stars with circumstellar disks (Class
II), reflecting a rapid evolutionary phase between these stages
\citep{{Kenyon1995}, {Wood2002}}. The presence of
comparable distribution features in the Cloud 2 clusters suggests that
similar evolutionary processes are at play.
In addition, the peak around $\alpha \sim 1.0$ corresponds to Class I
sources. Class I objects represent an early protostellar phase,
characterized by active accretion and the presence of substantial
envelope material.
The significant number of Class I sources observed in the target
clusters strongly supports the interpretation that these regions are
in a very early stage of star formation.

In the target clusters analyzed in this study, the fraction of Class I
objects was found to be
17\% and 28\% for Cloud 2-N and -S clusters, respectively.
{In nearby star-forming regions,} the typical duration of the
Class I phase has been estimated to be about 0.5 Myr based on the
relation $N_{\rm Class I}/N_{\rm Class II} * \tau_{\rm disk}$
\citep{{Evans2009},{Dunham2015}}, and it is well established that the
Class I fraction decreases consistently with age. Specifically, the
Class I fraction is over 20\% for ages less than 0.5 Myr, rapidly
decreases to less than 10\% at around 1 Myr, and becomes almost 0\%
after 2 Myr. This trend supports the short duration of the
protostellar phase ($\lesssim$0.5 Myr) and the rapid stellar evolution
observed. This age-dependent decline is supported by the following
observational results: at ages younger than 1 Myr (Serpens, $\rho$
Ophiuchi), the Class I fraction is about 20--30\% \citep{Evans2009};
at 1 Myr (Orion), it drops to $\sim$5--10\% \citep{Megeath2012}; at
1--2 Myr (Taurus), it remains at $\sim$2--5\% \citep{Luhman2010}; at
2--3 Myr (IC 348), it further decreases to $<$1\%
(\citealt{{Lada2006},{Muench2007}}),
and at 5 Myr (NGC 2362), it becomes less than 1\% \citep{Dahm2007}.
Given that the target clusters are known to be very young ($<$1 Myr),
the relatively high Class I fraction is consistent with the trend for
nearby star-forming regions.

The fractions of evolved disks in Cloud 2-N and Cloud 2-S were found
to be 18\% and 11\%, respectively.  The
fraction of evolved disks and its dependence on age have been observed
across various star-forming regions:
\citet{Luhman2010} reported that in the Taurus region, approximately
15\% of K5--M5 type stars possess evolved or transitional disks. Based
on this fraction, the duration of the evolved disk phase was estimated
to be about 0.45 Myr. Furthermore, a meta-analysis by
\citet{Ribas2014}, which compiled data from multiple star-forming
regions, revealed that the fraction of evolved disks changes with age
as follows: at 1 Myr, the fraction is less than 10\%, with primordial
disks being dominant; at 5 Myr, the fraction of evolved disks
increases to around 30\%, while the proportion of optically thick
disks decreases; and by 10 Myr, the evolved disk fraction also
declines to about 10\%, with diskless stars becoming the dominant
population. This trend has been interpreted as consistent with the
``inside-out clearing'' model, in which disk dissipation progresses
from the inner regions outward \citep{Ribas2014}.
While the observed fraction of evolved disks in the Cloud 2 clusters
is slightly higher, it generally follows the trends seen in nearby
star-forming regions of similar ages.

{The differences in the Class I and evolved-disk fractions between
  Cloud 2-N and Cloud 2-S (17\% vs. 28\% for Class I; 18\% vs. 11\%
  for evolved-disk sources) may suggest that Cloud 2-S is younger,
  although the uncertainties in these fractions are likely
  large. Nevertheless, this interpretation is consistent with other
  indicators, such as the higher stellar density of Cloud 2-S (see
  Section~\ref{sec:distribution}).}

\subsubsection{Mass Dependence of Disk Properties}
\label{sec:mass_dependence}

To investigate whether the evolutionary stage of circumstellar disks
depends on stellar mass, we examined the relationship between the SED
slope $\alpha$ and the {dereddened F200W magnitude $(m_{\rm
    F200W})_0$,} which serves as a proxy for stellar mass.
{The intrinsic magnitude $(m_{\rm F200W})_0$ was obtained by
  estimating individual extinction values $A_V$ for each source,
  tracing back along the reddening vector to the 0.1 Myr isochrone in
  the $(m_{\rm F115W} - m_{\rm F200W})$ vs. $m_{\rm F200W}$ diagram
  (Figure~\ref{fig:cc_nircams}).}

According to {the 0.1 Myr isochrone track,} {dereddened F200W
  magnitudes between 14 to 18 mag correspond to stellar masses from
  $\sim$1.0 $M_\odot$ down to $\sim$0.1 $M_\odot$, while magnitudes
  between 18 to 20 mag correspond to the substellar regime down to
  $\sim$0.04 $M_\odot$.
While the sampled mass range is relatively narrow, it still allows us
to examine potential trends in disk evolution across a range of
stellar masses.}

Figure~\ref{fig:sed_f200w} plots {$\alpha$ against $(m_{\rm
    F200W})_0$} for sources in the Mass-$A_V$-limited sample defined
in Section~\ref{sec:dust_diskfraction}.
No clear trend or correlation is observed, suggesting that the SED
slope---and thus the disk evolutionary stage---is not strongly
dependent on stellar mass within the observed range. This implies that
disk evolution proceeds similarly across the sampled mass range, at
least above the detection limit.

To maintain clarity, only the Mass-$A_V$-limited sample is shown in
Figure~\ref{fig:sed_f200w}. Similar distributions are seen in the full
and $A_V$-limited samples, but including multiple samples with
different symbols would overcrowd the plot and obscure any potential
trends.

As discussed in Section~\ref{sec:sed_dist} and illustrated in
Figure~\ref{fig:sed_dist}, Class I sources are largely absent from the
Mass-$A_V$-limited sample due to their typical non-detection in the
short-wavelength NIR bands (F150W and F200W) used for mass
estimation. Even when detected, their brightness is likely dominated
by disk excess emission rather than photospheric flux, making the
observed magnitude an unreliable indicator of stellar mass. Therefore,
Class I sources are excluded from this analysis, as their inclusion
would not yield meaningful insights into the mass dependence of disk
properties.

\subsubsection{Spatial Distribution of disk-classified sources} 
\label{sec:distribution}

The spatial distribution of sources classified by disk type within the
cluster region is shown in Figure~\ref{fig:distribution}.
The left and right panels correspond to the Cloud 2-N
and Cloud 2-S clusters, respectively.
The bottom-left panels present the positions of individual sources.
In these panels, only sources belonging to the Mass–$A_V$–limited
sample and Class I sources are plotted. While
Section~\ref{sec:mass_dependence} focused exclusively on the
Mass–$A_V$–limited sample, here we additionally include Class I
sources to provide a more complete view of the spatial distribution of
disk-bearing objects. Both groups are shown in black, using the same
symbols as in Figure~\ref{fig:CM_MIRI_NIRCam}.

More than half of Class I sources are excluded from the
Mass–$A_V$–limited sample, as they are typically undetected in
short-wavelength NIR bands. However, as noted in
Section~\ref{sec:sed_dist}, Class I sources located within the cluster
regions are considered highly probable cluster members based on their
spatial concentration and the rarity of such objects. Therefore, they
are included in the plot to complement the Mass–$A_V$–limited sample
and better represent the overall disk population.

The top-left and bottom-right panels of Figure~\ref{fig:distribution}
present the fraction of optically thick disks as a function of
position along the RA (east–west) and Dec (north–south) directions,
respectively. Histograms represent the disk fractions within each
spatial bin, and error bars indicate Poisson uncertainties. All
statistical analyses here are based on the Mass–$A_V$–limited sample,
as defined in Section~\ref{sec:dust_diskfraction}.

These plots reveal no significant spatial variation in the
distribution of optically thick disks. Specifically, the disk
fractions remain relatively constant across both RA and Dec,
suggesting a lack of positional dependence. The bottom-left panel of
Figure~\ref{fig:distribution} shows the spatial distribution of
sources identified as having gas disks in Section 4.2, marked with
cyan rectangles. Although the sample size is small and no detailed
statistical analysis is performed, no apparent spatial trend is
observed.

These plots reveal no significant spatial variation in the distribution
of optically thick disks. Specifically, the disk fractions remain
relatively constant across both RA and Dec, suggesting a lack of
positional dependence. Sources identified as having gas disks in
Section~\ref{sec:gas_disk} are also marked with cyan rectangles in the
bottom-left panel.
Although the sample size is small and no detailed statistical analysis
is performed, no apparent spatial trend is observed.

In other star-forming regions, strong UV radiation from nearby
early-type stars (within $\sim$1 pc) has been proposed as a mechanism
for external disk dispersal (e.g., \citealt{Winter2022}). However, the
clusters studied here do not contain any early-type stars (see
\citealt{Yasui2024}). While \citealt{de Geus1993} reported the presence
of an \ion{H}{2} region in the broader Digel Cloud 2 area based on
H$\alpha$ observations, this region is not associated with the clusters
analyzed in this study. Instead, it is believed to have been ionized by
an older stellar population (e.g., MR-1 or IRS 1; see
\citealt{{Kobayashi2000}, {Kobayashi2008}}), likely formed one to two
generations earlier.

Therefore, the absence of any clear spatial trends in disk
classification is consistent with the current environment, which lacks
strong external UV sources capable of influencing disk evolution.
{Although no clear spatial trend in disk fraction is observed, the
difference in stellar density between Cloud 2-N and Cloud 2-S may
provide an additional clue to their evolutionary status. Cloud 2-S
appears more compact and has a higher stellar density, which together
may suggest a younger evolutionary stage compared to Cloud 2-N, unless
the two clusters are at very different distance.
Assuming a typical velocity dispersion of 1\,km\,s$^{-1}$, the
dynamical timescale for Cloud 2-S (radius $\approx$12$\arcsec$ at a
distance of 8 kpc) and Cloud 2-N (radius $\approx$15$\arcsec$ at the
same distance) is $4.6\times 10^5$ yr and $5.7\times 10^5$ yr,
respectively (radii from \citealt{Yasui2024}).  These values are
consistent with the expected order of magnitude for such clusters. We
consider this as a potential implication rather than a definitive
conclusion, given the large uncertainties involved.}

\subsection{Disk evolution in a low-metallicity environment}
\label{sec:diskevolve_lowmeta}

In Section~\ref{sec:dust_disk}, it was revealed that a large fraction
of sources in the Cloud 2 cluster exhibit NIR/MIR color excess at
wavelengths $\ge$3 $\mu$m, indicating that they retain dusty disks at
a high fraction, similar to nearby young star-forming clusters of
comparable age. On the other hand, the innermost regions of the disks,
traced by the 2 $\mu$m band, appear to have already dissipated. This
trend is not commonly observed in nearby young star-forming regions.
In Section~\ref{sec:gas_disk}, excess detected using a narrow-band
filter that covers the hydrogen emission line Br$\alpha$ suggests that
$\sim$35\% of the sources in the Cloud 2 clusters exhibit relatively
high mass accretion rates. This implies that at least the inner gas
disks are still present. Therefore, it is considered that only the
innermost dusty disks have dissipated or are not detectable within the
sensitivity and wavelength range of our observations.
Furthermore, in Section~\ref{sec:distribution}, it was shown that the
above-mentioned trends in both dusty and gaseous disks do not exhibit
any significant spatial dependence within the cluster.

Based on these observational results, we examine the disk evolution of
sources in the Cloud 2 clusters under the low-metallicity environment
in the outer Galaxy. In particular, for objects that do not show color
excess at 2 $\mu$m, we classify the possible physical causes into the
following four categories and discuss the likelihood of each. This
discussion focuses on objects with masses down to approximately 0.1
$M_\odot$, which are detected at wavelengths including the 7.7 $\mu$m
band and whose disk properties can be evaluated through their MIR
SEDs.
Note that objects in the brown dwarf mass regime ($<$0.1 $M_\odot$)
are known not to exhibit 2 $\mu$m excess \citep{Liu2003}, and their
protoplanetary disks will be discussed separately in
Section~\ref{sec:disk_bd}.

\begin{enumerate}
    
\item Primordial absence of inner disks

A straightforward interpretation is that this may be attributed to an observational bias, whereby the lower dust content expected in low-metallicity environments leads to reduced opacity. This, in turn, weakens infrared emission at shorter wavelengths such as 2 $\mu$m in the NIR, making color excess more difficult to detect. However, classical disk models, such as the Minimum Mass Solar Nebula (MMSN), predict very high opacities at 0.1 au under solar metallicity conditions, with values on the order of $10^5$ (see details in \citealt{Yasui2009} and references therein). 
Therefore, it is unlikely that a decrease in metallicity by an order of magnitude would be sufficient to cause a transition from an optically thick to an optically thin state, although some reduction in opacity may occur.
Moreover, since disk models generally predict that gas surface density
decreases toward the outer regions of the disk, it is difficult to
attribute the non-detection of 2 $\mu$m excess solely to opacity effects
in low-metallicity environments.

\item Inner hole or cavity

Another possible explanation for the lack of 2 $\mu$m excess is the
presence of an inner hole or cavity in the disk. In this context, the
composition and thermal properties of dust grains may play a key
role. Dust is broadly classified into silicate-based (Si-based) and
carbonaceous (C-based) types, with condensation temperatures of
approximately 1000 K and 2000 K, respectively. The absence of infrared
excess at 2 $\mu$m in this study suggests that Si-based dust may be
dominant in this environment. Si-based grains are known to form under
the influence of supernova remnants (SNRs), which is consistent with
the environmental conditions of Cloud 2.

However, when applying theoretical models of dust formation in AGB and
SAGB stars, \citet{Ventura2012} indicate that the production of
Si-based dust increases with metallicity.
Observationally, a higher fraction of carbon stars has been reported
in low-metallicity environments such as the LMC, SMC, and the outer
regions of the Galaxy, which appears to contradict the interpretation
proposed here.

Cloud 2 may indeed exhibit dust properties that differ from those in
nearby star-forming regions, as suggested by the following
observational characteristics.
Cloud 2 is a very young star-forming region with a large number of
Class I objects, yet many of them are not detected in NIR observations
up to 2 $\mu$m, unlike nearby young star-forming regions such as
$\rho$ Ophiuchi, NGC 2024, and NGC 1333 (all younger than 1 Myr; e.g.,
\citealt{Wilking2008}).
In addition, Cloud 2 shows relatively weak extinction ($A_V \lesssim
10$ mag), which may be due to low-metallicity environments where a
lower dust-to-gas ratio is expected, whereas nearby regions exhibit
much stronger extinction in the infrared, with $A_V$ values of $>$10
mag (e.g., \citealt{Robberto2024}). 
These two observational characteristics support the possibility that
      dust properties in low-metallicity environments differ from those
      in the solar neighborhood.

\item Grain growth and settling

      When dust grains grow beyond micron sizes, their emission
      efficiency at shorter wavelengths decreases. In addition, as the
      grains settle toward the midplane of the disk, the surface
      emission diminishes, making excess emission around 2 $\mu$m less
      detectable. Dust growth is generally more efficient in the inner
      regions of disks, where the gas density is higher and the orbital
      period is shorter. In low-metallicity environments, the
      dust-to-gas mass ratio
      is lower, which reduces the frequency of collisions between
      particles and makes grain growth less efficient. As a result, the
      formation of large grains is suppressed, and the efficiency of
      dust settling is also expected to decrease \citep{Matsukoba2024}.
      Therefore, it is unlikely that dust growth and settling in the
      inner disk, as traced by 2 $\mu$m emission, would proceed more
      rapidly in low-metallicity environments than in solar-metallicity
      ones. (See also \citealt{Yasui2009} for further details.)

\item Inside-out disk dispersal

  In the solar neighborhood, it is well established that disk
  dispersal proceeds in an inside-out manner, where the inner disk is
  first dissipated by mass accretion, followed by the dissipation of
  the outer disk.  However, the difference in timescales between the
  dispersal of the inner and outer disks is considered to be very
  small (e.g., \citealt{Alexander2014}).
  In this context, we consider whether the Cloud 2 clusters might be
  in an intermediate evolutionary stage, where the inner disk has
  already dissipated while the outer disk remains detectable.

In low-metallicity environments, the dust-to-gas mass ratio is
expected to be lower, which may lead to a higher ionization degree in
the disk compared to solar-metallicity environments. As a result, both
mass accretion and disk winds could potentially operate more
efficiently \citep[e.g.,][]{{Ercolano2010}, {Bai2017},
  {Nakatani2018a}, {Nakatani2018b}}.
With regard to mass accretion, as shown in Section~\ref{sec:gas_disk},
relatively high accretion rates are suggested for Cloud 2 cluster
members. This indicates that accretion is still ongoing, implying the
continued presence of an inner gas disk. Therefore, it is unlikely
that the inner disk has already dissipated solely due to accretion.

With regard to disk winds, two major mechanisms have been proposed:
magnetohydrodynamic (MHD) disk winds and photoevaporative winds (e.g.,
\citealt{Pascucci2023}).  MHD disk winds are thought to work primarily
in the inner disk regions ($R_G \simeq 0.1$--1 au).  However,
regardless of whether this mechanism is more efficient in
low-metallicity environments, the presence of an inner gas disk in
Cloud 2 suggests that the inner disk has not yet been removed by this
process.
On the other hand, photoevaporative winds are expected to work in the
outer disk regions (e.g., $R_G \gtrsim 5$ au). In Cloud 2, color
excesses at wavelengths $\ge$3 $\mu$m are observed, indicating the
presence of outer dusty disks.
This suggests that the outer disk has also not yet been dissipated by
photoevaporation.

Taken together, these observations do not support the inside-out disk
dispersal scenario in Cloud 2. Both the inner and outer disk
components appear to be present, making this mechanism an unlikely
explanation for the observed lack of 2 $\mu$m excess.
    
\end{enumerate}

Among the four scenarios considered here, three---(1) primordial
absence of inner disks, (3) grain growth and settling, and (4)
inside-out disk dispersal---are inconsistent with the current
observational results and with commonly accepted physical
interpretations. These scenarios are therefore unlikely to explain the
observed lack of 2 $\mu$m excess. In contrast, although not yet fully
understood, the scenario that remains most consistent with the
observations is (2) the presence of an inner hole or cavity, which may
be attributed to dust properties specific to low-metallicity
environments.

\subsection{Disk Fraction of Brown Dwarfs Based on JWST/NIRCam Data}
\label{sec:disk_bd}

{In this section, we focus exclusively on brown dwarfs and derive
  their disk fraction using JWST/NIRCam photometry ($\le$4.4 $\mu$m).
The imaging sensitivity of JWST decreases toward longer wavelengths;
in particular, the photometric depth achievable with MIRI
($>$5\,$\mu$m) is substantially lower than that attained with NIRCam
at wavelengths below 5\,$\mu$m.
In contrast, data up to the NIRCam F444W band reach the substellar
regime needed to identify brown dwarfs (Section~\ref{sec:cm}).
Accordingly, we use NIRCam data up to F444W to assess the disk
properties of brown dwarfs, enabling a uniform and sensitive
determination of their disk fraction within the substellar regime.}

Because brown dwarfs are low-temperature objects, and at wavelengths
of $\lesssim$3 $\mu$m the emitting regions of their disks are very
small, it is difficult to detect disks at such wavelengths (e.g.,
\citealt{Liu2003}). Previous studies have mainly used wavelengths up
to the 5.8 $\mu$m band---one band longer than the 4.5 $\mu$m band in
Spitzer/IRAC—for detecting brown dwarf disks (e.g.,
\citealt{Luhman2005}, \citealt{Monin2010}, \citealt{Guieu2007},
\citealt{Riaz2009}).
However, {in the absence of JWST/MIRI F560W photometry
  here---which roughly corresponds to Spitzer's 5.8\,$\mu$m band---we
  adopt NIRCam-only diagnostics and assess disk presence using
  excesses measured up to the F444W band}
(see Appendix~\ref{sec_appendix:cc_dwarf} {for the evaluation of
  color--color combinations and the adopted criteria for disk
  identification).}

As a result, we found that brown dwarfs can be almost unambiguously
identified
{using the F444W-excess diagnostic diagrams shown in
  Figure~\ref{fig:CC_NIRCam_F444W}: ($m_{\rm F200W} - m_{\rm F444W}$)
  vs. ($m_{\rm F115W} - m_{\rm F150W}$) color--color diagrams.
Dwarf tracks are shown as blue curves.}
The boundary line indicating the presence or absence of a disk is
shown as a gray dot-dashed line, originating from the 0.1 $M_\odot$
point on the {dwarf track} and running parallel to the reddening
vector.
{In Figure~\ref{fig:CC_NIRCam_F444W}, sources detected up to the
  F770W band are plotted using the same symbol as in
  Figure~\ref{fig:CM_MIRI_NIRCam}.  Sources not detected in the F770W
  band are highlighted separately and are shown as red open squares.}
Those located to the right of the boundary can be considered to have
disks.

{To define a mass-limited sample appropriate for assessing
  brown-dwarf disk properties, we first examine the NIRCam SW
  color--magnitude diagrams ($m_{\rm F115W} - m_{\rm F200W}$
  vs. $m_{\rm F200W}$; Figure~\ref{fig:cc_nircam_bd}).
  Sources that lack F770W detections but are detected in the NIRCam
  bands are indicated by square symbols in
  Figure~\ref{fig:cc_nircam_bd}.}
In Figure~\ref{fig:cc_nircam_bd}, sources are color-coded based on
whether they belong to the $A_V$-limited sample defined in
Section~\ref{sec:ident_cluster}: black for $A_V$-limited and gray for
non-$A_V$-limited.
For the purpose of this study, we define a brown dwarf (BD)-limited
sample
{to establish a mass-limited selection that is consistent with the
  NIRCam-based detection limits. Although the physical
  hydrogen-burning limit is located at $\simeq$0.08 $M_\odot$,
  photometric mass estimates at this level are subject to substantial
  uncertainties due to age spread within the cluster, extinction
  variations, and model dependencies. We therefore adopt an
  observationally motivated boundary defined by the faintest detected
  Class III/EV source in the ($m_{\rm F150W} - m_{\rm F200W}$)
  vs. $m_{\rm F200W}$ diagram, corresponding to a mass of $\simeq$0.04
  $M_\odot$ (Section~\ref{sec:ident_cluster}.)}

{As a consequence of this definition, sources not detected in the
  F770W band are found predominantly at or below this mass threshold
  (cf. Figure~\ref{fig:cc_nircam_bd}). These sources are therefore
  included in the disk fraction analysis only when their disk
  properties can be reliably classified using the NIRCam F444W-excess
  diagnostics (see Appendix~\ref{sec_appendix:cc_dwarf}).  The red
  line in Figures~\ref{fig:cc_nircams} and \ref{fig:cc_nircam_bd}
  marks this boundary. By combining this BD-limited sample with the
  $A_V$-limited classification, we define a ``BD-$A_V$-limited
  sample.''}

In the Cloud 2-N cluster, Figure~\ref{fig:cc_nircam_bd} shows that
sources not detected in the F770W band but detected up to the F444W
band are {predominantly} considered to have {masses below the
  BD-limited boundary ($\simeq$0.04 $M_\odot$).} {At the same
  time,} Figure~\ref{fig:cc_nircams} indicates that even among sources
detected in the F770W band, some also fall {within the BD-limited
  range.}
{Although F770W has the lowest spatial resolution among the bands
  used here, its detection completeness remains sufficiently high (see
  Section~\ref{sec:dust_diskfraction}) to enable reliable detections
  well into the substellar regime.}

{Consistent with the BD-limited definition above, candidate brown
  dwarfs are defined as sources that lie within the BD–$A_V$-limited
  region and are detected at least up to the F444W band. Disk presence
  is then evaluated using two complementary diagnostics: for sources
  detected at F770W, MIR SED slopes are used; for sources not detected
  at F770W but detected at F444W, F444W-excess diagrams
  (Figure~\ref{fig:CC_NIRCam_F444W}) are adopted. Sources undetected
  in F444W lack sufficient information for disk classification and are
  therefore excluded from the disk/no-disk assessment.}

Among the sources detected in the F770W band, seven objects are
identified: one Class I, five Class II, and one EV source.
As in the derivation of the stellar disk fraction
(Section~\ref{sec:dust_diskfraction}), the Class I source is excluded
from the disk fraction calculation.
Additionally, 10 sources are detected only up to the F444W band, with
four classified as having disks and five as diskless. Considering all
these sources, excluding the Class I object, 10 out of 15 brown dwarf
candidates are identified as having disks, resulting in a brown dwarf
disk fraction of 67$\pm$27\% (10/15) in the Cloud 2-N cluster.

In the Cloud 2-S cluster, Figure~\ref{fig:cc_nircam_bd} shows that
sources not detected in the F770W band but detected up to the F444W
band include objects with masses {both} above and below {the
  BD limited boundary}. This contrasts with the Cloud 2-N cluster,
where such sources are almost exclusively substellar.
As discussed in Section~\ref{sec:dust_diskfraction}, the completeness
of F770W detections in Cloud 2-S is affected by observational
limitations, particularly source confusion arising from the higher
stellar density in this region and the larger FWHM of the F770W band
compared to F444W. The mass distribution of the F770W non-detected
sources in Figure~\ref{fig:cc_nircam_bd} is consistent with this
picture, rather than being driven solely by intrinsic faintness.
Meanwhile, Figure~\ref{fig:cc_nircams} indicates that, similar to
Cloud 2-N, some sources detected in the F770W band in Cloud 2-S also
fall below the BD-limited boundary, suggesting the presence of brown
dwarfs among them.

To estimate the brown dwarf disk fraction in Cloud 2-S, we follow the
same approach as in Cloud 2-N.
Among the sources detected in the F770W band and included in the
BD-$A_V$-limited sample, seven objects are identified: six Class II
sources and one EV sources.
As in previous analyses, the Class I sources are excluded from the
disk fraction calculation.
In addition, among sources detected only up to the F444W band within
the BD-$A_V$-limited sample, 18 are classified as having disks and six
as diskless.
Combining these results, we find that out of 31 brown dwarf
candidates, 24 are classified as having disks, yielding a brown dwarf
disk fraction of 77$\pm$21\% (24/31).

For completeness and to test the robustness of our results against the
choice of the stellar/brown-dwarf boundary, we also compute
brown-dwarf disk fractions adopting the canonical threshold of
0.08\,$M_\odot$ (dotted red line in Figures~\ref{fig:cc_nircams} and
\ref{fig:cc_nircam_bd}). Using this boundary, the disk fractions are
61$\pm$17\% (20/33) for Cloud 2-N and 86$\pm$27\% (19/22) for Cloud
2-S, which are identical within rounding uncertainties to the values
derived using the operational $\simeq$0.04 $M_\odot$ BD-limited
boundary (solid red line).
This invariance indicates that sources in the mass range between 0.04
and 0.08 $M_\odot$ contribute proportionally to the disk-bearing and
diskless populations.
We caution that, as discussed above, age spreads among cluster members
and the reliance on photometric mass estimates limit the physical
interpretation of any strict mass cutoff.

The brown dwarf disk fractions derived for the Cloud 2 clusters 
are comparable to the stellar disk fraction ($\sim$75\%) derived from
MIR SED slopes up to 7.7 $\mu$m. These results suggest that, in these
very young clusters, brown dwarfs retain their disks at a rate similar
to that of stellar-mass objects.

Previous studies in the solar neighborhood have shown that disk
evolution for brown dwarfs proceeds on a timescale comparable to that
of stars (e.g., \citealt{Luhman2012}). Therefore, in very young
clusters such as those studied here, a relatively high disk fraction
is expected if disk evolution follows similar trends across different
mass regimes.
The results here indicate that this trend also holds in the
low-metallicity environment of the Cloud 2 clusters. Brown dwarfs,
like their stellar counterparts, exhibit high disk presence at this
early evolutionary stage.
This suggests that, at least during the initial phases of disk
evolution, the presence and early evolution of primordial disks are
not strongly dependent on the mass of the central object, whether
stellar or substellar.

Although previous studies have reported that the IMF peak mass (i.e.,
characteristic mass) in these clusters is lower than that in the solar
neighborhood \citep{Yasui2024}, implying a relative abundance of brown
dwarf–mass objects, the disk fractions observed here do not show
significant deviation due to metallicity. Thus, despite the unique
mass distribution, the disk evolution behavior appears to be broadly
consistent with that observed in higher-metallicity environments.

\section{Conclusion} \label{sec:conclusion}

This study aims to elucidate the evolution of protoplanetary disks in
environments with low metallicity, where low dust content is also
expected, conditions resembling those of the early universe.  Digel
Cloud 2, located in the outer Galaxy (distance $D \simeq 8$ kpc), is
known for its significantly lower metallicity compared to the solar
neighborhood ($\simeq$$-$0.7 dex), making it an ideal target for
investigating planet formation in primordial
environments. Understanding disk evolution in such low-metallicity
environments is crucial for testing the universality of planet
formation.

To this end, we conducted imaging observations of two clusters in
Digel Cloud 2 (Cloud 2-N and Cloud 2-S), a very young ($\simeq$0.1
Myr) star-forming region, using JWST/NIRCam and MIRI across the 1--20
$\mu$m wavelength range. In this paper, we focus on four bands, F356W,
F444W, F405N, and F770W, to capture disk properties from inner to
outer regions in the MIR. The achieved spatial resolution and
sensitivity are comparable to Spitzer observations in the solar
neighborhood ($\sim$500 pc), providing the first observational
framework for directly comparing disk evolution in the outer Galaxy
with nearby regions.

Leveraging JWST's capabilities, we capture the evolutionary stages of
protoplanetary disks in detail, especially using the F770W band. While
NIR observations ($\lesssim$2 $\mu$m) are limited to probing only the
innermost disk regions ($\lesssim$0.1 au), {our study enables}
detection of more extended regions ($\sim$1 au). Moreover, F770W
allows detection of both optically thick and thin disks, enabling a
more comprehensive understanding of disk properties.

\begin{enumerate}
\item In the F770W band, a detection limit of $\simeq$0.1 $M_\odot$
  and spatial resolution of $\lesssim$$0\farcs3$ are achieved, resulting
  in the detection of 89 and 95 sources in Cloud 2-N and Cloud
  2-S, respectively.
  In Cloud 2-N, nearly all sources with masses $\gtrsim$0.1 $M_\odot$
  detected in shorter-wavelength bands (F356W, F444W) were also
  detected in F770W, indicating high completeness for stellar-mass
  objects.
  In contrast, Cloud 2-S showed reduced completeness in F770W, likely
  due to source confusion in its higher-density environment.

\item MIR SED slopes ($\alpha$) were derived using F356W, F444W, and
  F770W data, allowing classification of disk evolutionary stages:
  Class I ($\alpha > 0$), Class II ($-1.8 < \alpha \leq 0$), evolved
  disks ($-2.56 \lesssim \alpha < -1.8$), and Class III ($\alpha <
  -2.56$).
  Among F770W-detected sources, Cloud 2-N contained 12 Class I, 50
  Class II, 14 evolved disk, and 13 Class III objects, while Cloud 2-S
  contained 23, 53, 13, and 6, respectively.
  To estimate disk fractions among stellar-mass cluster members, we
  defined a Mass-$A_V$-limited sample by selecting sources with masses
  $\gtrsim$0.1 $M_\odot$ and extinctions consistent with the cluster
  environment.  The mass threshold ensures that the sample includes
  only stellar-mass objects, while the extinction criterion helps to
  exclude foreground stars and isolate likely cluster members.  Based
  on this sample, the fractions of optically thick disks (Class II)
  are estimated at 74$\pm$15 \% in Cloud 2-N and 81$\pm$17 \% in Cloud
  2-S.
These results suggest that, at early evolutionary stages, a large fraction of objects in low-metallicity environments also retain optically thick disks, similar to those observed in the solar neighborhood.

\item In the Cloud 2 clusters, disk evolution trends based on infrared
  excess at different wavelengths show that 3--4 $\mu$m excess aligns
  well with MIR SED-based classifications up to 7.7 $\mu$m, reflecting
  dust disk properties.  However, a large fraction of Class I and
  Class II objects lack excess emission at 2 $\mu$m, suggesting
  reduced emission from the innermost, hottest disk material.
  This trend contrasts with the $\sim$75\% disk fraction
  observed across the 2--7.7 $\mu$m range in nearby young star-forming
  regions, indicating that the 2 $\mu$m band alone reveals a distinct
  disk evolution pattern in the Cloud 2 environment.

 \item In addition to probing dust disk classifications through
   broad-band MIR observations, mass accretion activity is
   investigated using the F405N band, which covers the Br$\alpha$
   line. This narrow-band filter {achieves} a detection limit of
   $\simeq$0.1 $M_\odot$, comparable to that of F770W.
   We define a criterion for identifying mass-accreting sources using
   color--color diagrams based on this band, classifying objects with
   Br$\alpha$ emission equivalent widths greater than 50 \AA\ as
   accreting sources.
   Based on this method, $\sim$35\% of sources in the Cloud 2 clusters
   are identified as accreting sources, primarily among Class I and II
   objects.
   These accretion rates correspond to Br$\alpha$ equivalent widths
   exceeding the 50 \AA\ threshold, and are estimated to be
   $\gtrsim$$10^{-6}$ $M_\odot$~yr$^{-1}$. They are comparable to or
   even higher than those observed in similarly aged low-mass stars in
   the solar neighborhood, suggesting that a sufficient amount of gas
   remains in the inner disks of Cloud 2 objects to sustain active
   accretion.

 \item To explain the observed lack of color excess specifically at
   NIR 2 $\mu$m in the Cloud 2 clusters, we considered four possible
   physical scenarios: (1) primordial absence of inner disk dust, (2)
   the presence of inner cavities, (3) grain growth and settling, and
   (4) inside-out disk dispersal. Among these, scenarios (1), (3), and
   (4) were found to be inconsistent with the current observational
   results and theoretical understanding. In contrast, the inner
   cavity hypothesis (2) remains the most plausible explanation, as it
   can be attributed to dust properties unique to low-metallicity
   environments.
   In particular, the notably low detection rate of Class I sources at
   2 $\mu$m and the relatively small extinction values observed in
   Cloud 2, compared to nearby young star-forming regions, may support
   the idea that the dust in this environment possesses different
   physical properties.

     \item To complement the analysis of low-mass stars down to
       0.1 $M_\odot$, we investigated the presence of optically thick
       disks around brown dwarf mass objects ($\simeq$0.01--0.1
       $M_\odot$) in the Cloud 2 clusters, using infrared excess in
       the F444W band as a diagnostic.
       Based on this method, the disk fraction among brown dwarf
       candidates in the Cloud 2 clusters is estimated to be
       $\sim$75\%, which is comparable to the stellar disk fraction
       derived from MIR SED slopes up to 7.7 $\mu$m. These results
       suggest that, in these very young clusters, brown dwarfs retain
       their disks at a rate similar to that of stellar-mass objects.
       This trend is consistent with previous findings in
       solar-metallicity environments, and the similar results
       obtained in the low-metallicity Cloud 2 clusters indicate that
       early disk evolution is not strongly dependent on the mass of
       the central object or on metallicity. Although the initial mass
       function (IMF) in this region suggests a somewhat different
       distribution, the disk retention rates appear broadly
       consistent across mass ranges.

\end{enumerate}


  
\section*{Acknowledgements}
  We thank Dr. Itsuki Sakon for helpful discussions on dust physics in
  low-metallicity environments. C.Y. acknowledges support from JSPS
  KAKENHI Grant Number 25K07363.
  The data presented in this article were obtained from the Mikulski
  Archive for Space Telescopes (MAST) at the Space Telescope Science
  Institute. The specific observations analyzed can be accessed via
  doi:10.17909/a2k0-tf62. 


%



\software{
  astropy \citep{{Astropy2013}, {Astropy2018}, {Astropy2022}}, 
photutils \citep{Bradley2022}. 
          }



\clearpage

\begin{figure*}
  \begin{center}
    \gridline{\fig{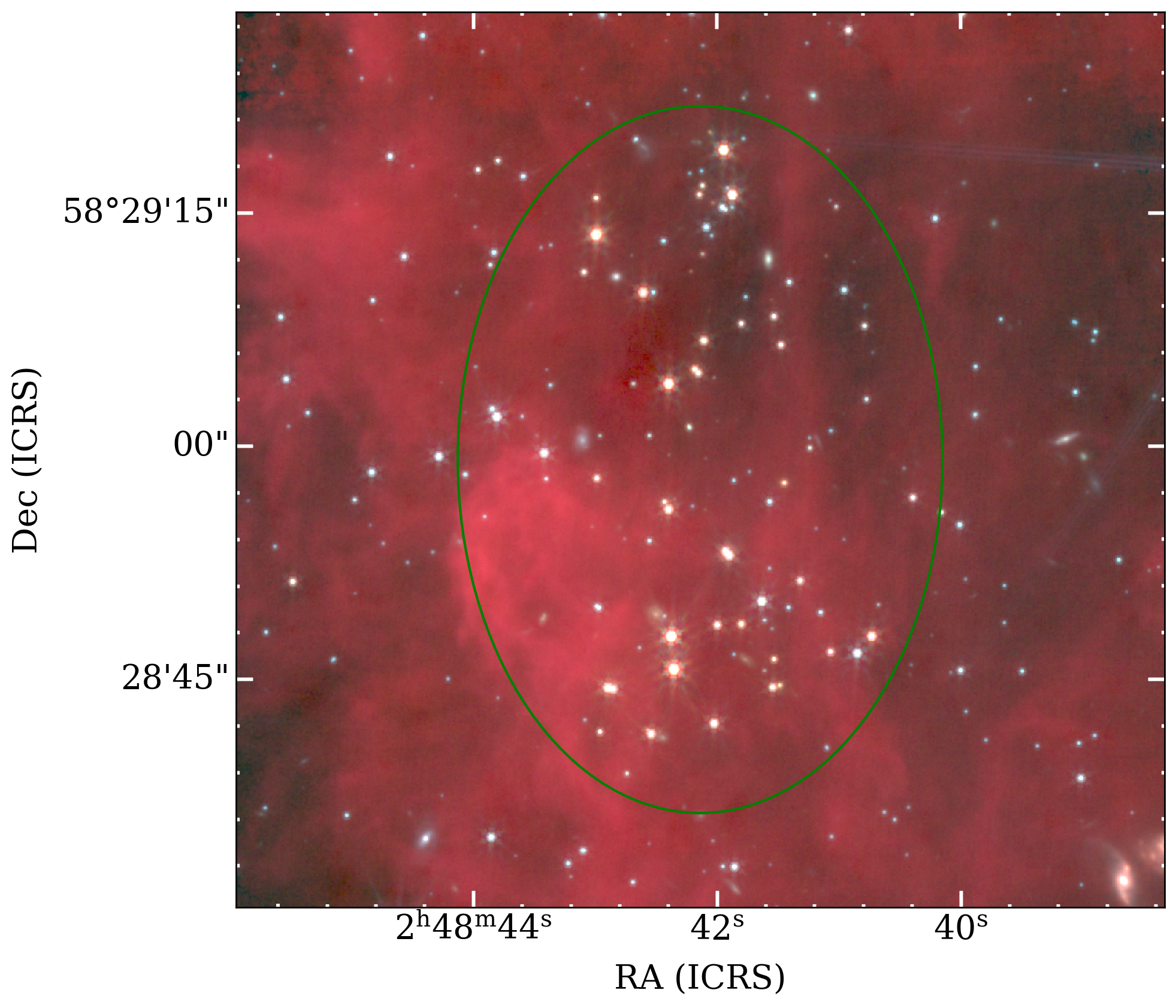}{0.6\textwidth}{}}
    \gridline{\fig{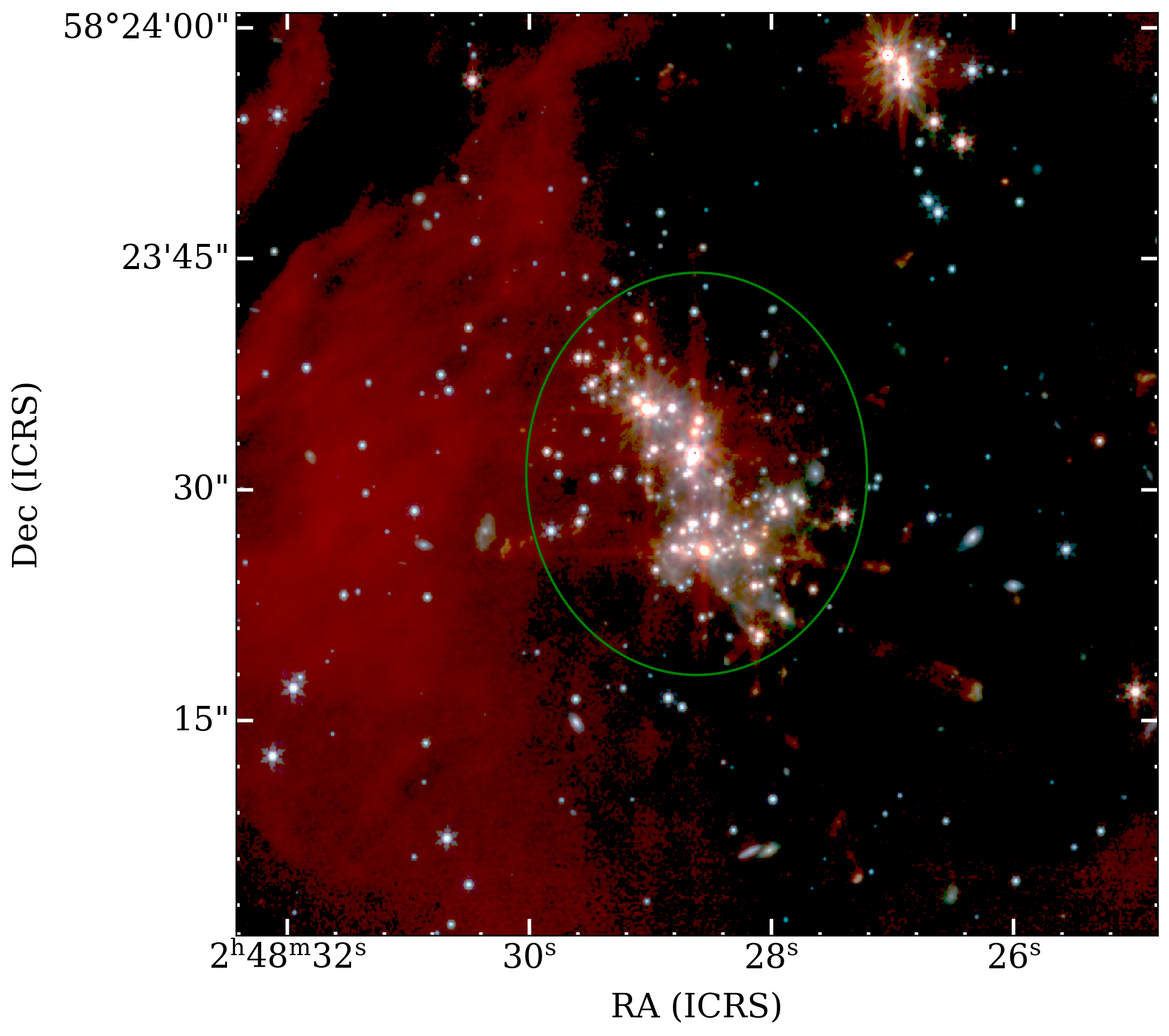}{0.6\textwidth}{}}
    \caption{Zoomed-in pseudo-color images of the Cloud 2-N and 2-S
      clusters obtained with JWST/NIRCam LW and MIRI, produced by
      combining the NIRCam images F356W (blue) and F444W (green), and
      the MIRI image F770W (red), representing the MIR band. North is
      up and east is to the left. The image areas, each with a field
      of view of approximately $1' \times 1'$, which correspond to the
      white squares shown in Figures~1 and 2 of \citet{Yasui2024}, and
      cover the same sky region as Figure~3 in the reference.
      The positions of the cluster regions are shown as green
      ellipses.}
\label{fig:CL2NScl}
\end{center}
\end{figure*}

\begin{figure}
\begin{center}
  \includegraphics[width=0.48\textwidth]{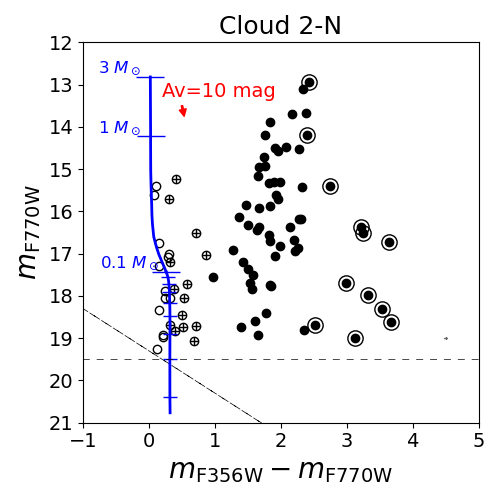}
    \hspace{0.1cm}  
  \includegraphics[width=0.48\textwidth]{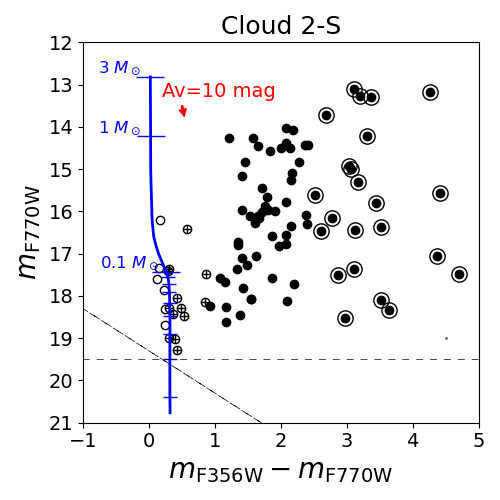}
 \caption{MIRI-NIRCam color--magnitude diagram, $(m_{\rm F356W} -
   m_{\rm F770W})$ vs. $m_{\rm F770W}$, for the Cloud 2-N and 2-S
   clusters.
   All sources detected with more than 10$\sigma$ in F356W and F770W
   bands are plotted.
   Class I, II, EV, and III objects are represented by black filled
   circles enclosed in circles, black filled circles, encircled plus
   symbols, and black open circles, respectively.
   The legend shown in the left panel applies to both panels and will
   be used consistently in subsequent figures.
   Isochrone tracks for the age of 0.1 Myr are shown with blue
     curves.
   Large tick marks on the isochrones denote masses of 3, 1, and 0.1
   $M_\odot$ (from top to bottom), and the smaller tick marks below
   them denote 0.09, 0.08, …, 0.02 $M_\odot$.
   The sensitivity limits, corresponding to the 10$\sigma$ limiting
   magnitudes of the F770W and F356W bands, are indicated by dashed
   lines. 
   Typical photometric uncertainties are indicated by thick gray error
   bars in the lower-right corners of both panels.
   The red arrows show the reddening vectors of $A_V = 10$ mag.}
\label{fig:CM_MIRI_NIRCam}
\end{center}
\end{figure}

\begin{figure}
\begin{center}
  \includegraphics[width=0.48\textwidth]{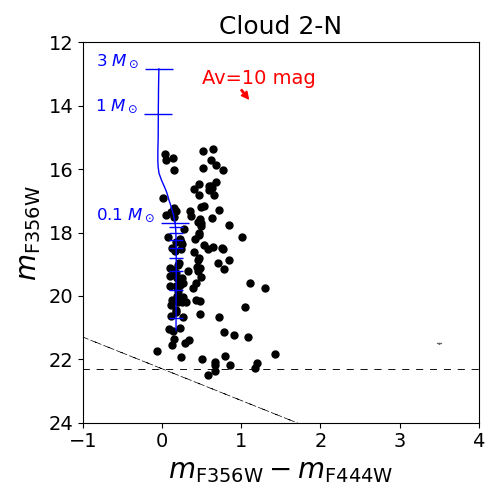}
  \includegraphics[width=0.48\textwidth]{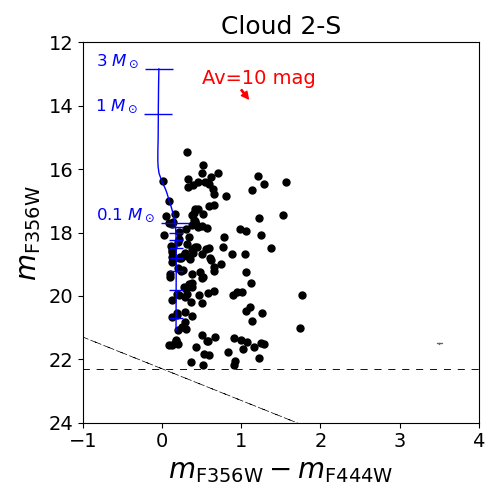}

 \caption{NIRCam LW color--magnitude diagram, $(m_{\rm F356W} -
   m_{\rm F444W})$ vs. $m_{\rm F356W}$, for the Cloud 2-N and 2-S
   clusters.
   All sources detected with more than 10$\sigma$ in both F356W and
   F444W bands are plotted.
   Isochrone tracks for the age of 0.1 Myr are shown with blue
   curves.
   Tick marks on the isochrones are the same as in
   Figure~\ref{fig:CM_MIRI_NIRCam}.
   The sensitivity limits, corresponding to the 10$\sigma$ limiting
   magnitudes of the F356W and F444W bands, are indicated by dashed
   lines. 
   Typical photometric uncertainties are indicated by thick gray error
   bars in the lower-right corners of both panels.
   The red arrows show the reddening vectors of $A_V = 10$ mag.}
\label{fig:CM_F356W_F444W}
\end{center}
\end{figure}

\begin{figure}
\begin{center}
  \includegraphics[width=0.48\textwidth]{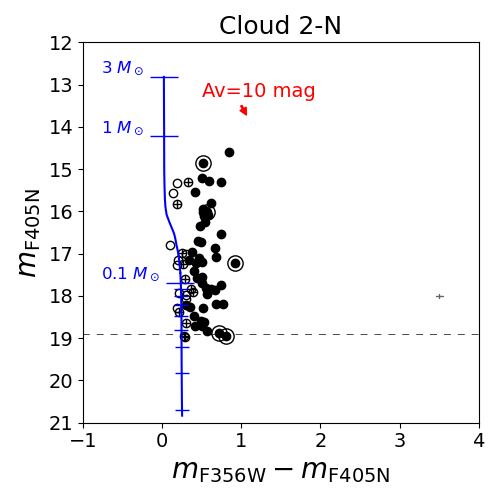}
  \includegraphics[width=0.48\textwidth]{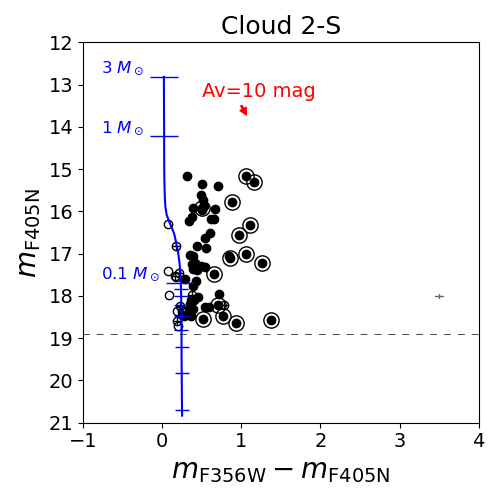}

 \caption{F405N color--magnitude diagram, $(m_{\rm F356W} - m_{\rm
     F405N})$ vs. $m_{\rm F405N}$, for the Cloud 2-N and 2-S clusters.
   All sources detected with more than 10$\sigma$ in both F356W and
   F405N bands are plotted.
   The symbols are the same as in Figure~\ref{fig:CM_MIRI_NIRCam},
   based on the classification from the MIR SED slope ($\alpha$).
   Isochrone tracks for the age of 0.1 Myr are shown with blue curves.
   Tick marks on isochrones are the same as in
   Figure~\ref{fig:CM_MIRI_NIRCam}.  The sensitivity limits,
   corresponding to the 10$\sigma$ limiting magnitudes of the F405N
   bands, are indicated by dashed lines.
   Typical photometric uncertainties are indicated by thick gray error
   bars in the lower-right corners of both panels.
   The red arrows show the reddening vectors of $A_V = 10$ mag.}
\label{fig:CM_F356W_F405N}
\end{center}
\end{figure}

\begin{figure}
\begin{center}
 \includegraphics[scale=0.7]{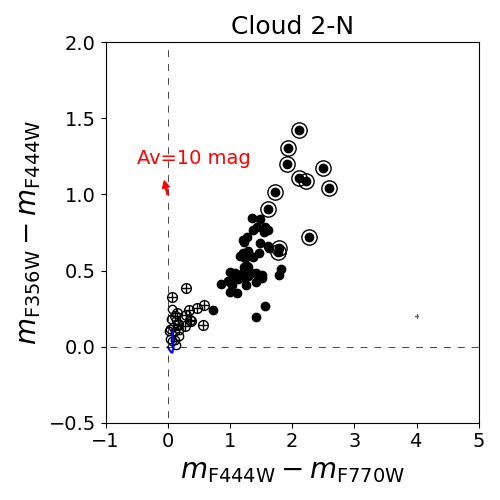}
 \includegraphics[scale=0.7]{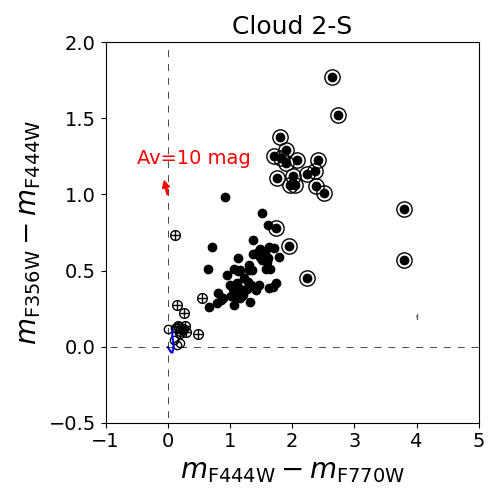}
 
 \caption{MIRI-NIRCam color--color diagram, $(m_{\rm F444W} - m_{\rm
     F770W})$ vs. $(m_{\rm F356W} - m_{\rm F444W})$, for the Cloud 2 clusters.
   The symbols are the same as in Figure~\ref{fig:CM_MIRI_NIRCam},
   based on the classification from the MIR SED slope ($\alpha$).
   The blue curve represents the dwarf track, defined as the locus of
   points corresponding to unreddened main-sequence stars.
   Typical photometric uncertainties are indicated by thick gray error
   bars in the lower-right corners of both panels.}
  \label{fig:CC_MIRI_NIRCam}
\end{center}
\end{figure}

\begin{figure}
\begin{center}
  \includegraphics[scale=0.7]{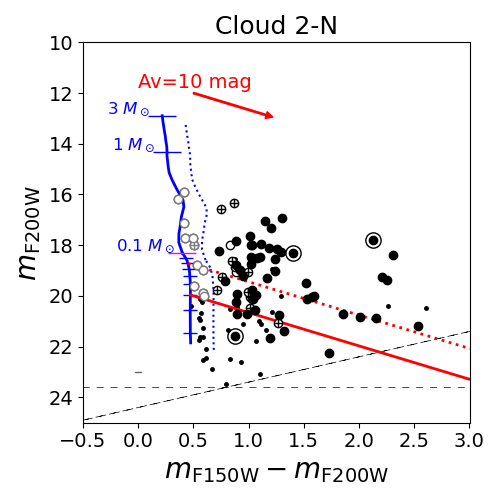}
 \includegraphics[scale=0.7]{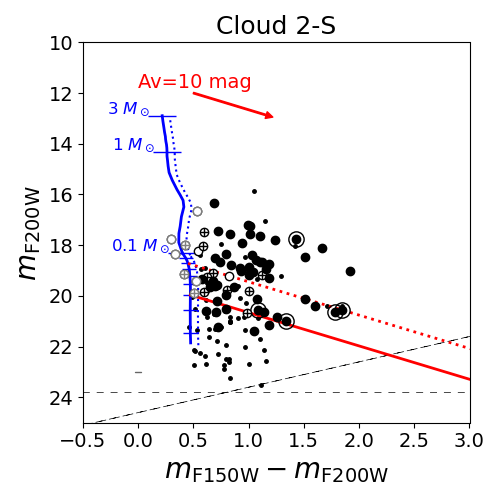}
  \caption{NIRCam SW color--magnitude diagram, $(m_{\rm F115W} -
    m_{\rm F200W})$ vs. $m_{\rm F200W}$, for the Cloud 2 cluster
    regions.
    Circles indicate sources for which the MIR SED slope ($\alpha$) is
    derived, with detection of F770W band.
    These use the same symbols as in Figure~\ref{fig:CM_MIRI_NIRCam},
    classified according to $\alpha$.
    Typical photometric uncertainties for these sources are indicated
    by thick gray error bars in the lower-left corner.
    In addition, small dots represent sources detected in the NIRCam
    bands but non-detected in the F770W band.
    Isochrone tracks for the age of 0.1 Myr are shown with blue
    curves.  Tick marks on the isochrones are the same as in
    Figure~\ref{fig:CM_MIRI_NIRCam}.
    The sensitivity limits, corresponding to the 10$\sigma$ limiting
    magnitudes of the F150W and F200W bands, are indicated by dashed
    lines.
    The extinction vector corresponding to $A_V=10$ mag is indicated
    by red arrows.
    The shifted track (dotted blue curves) is used as the boundary for
    selecting $A_V$-limited sample sources. Sources within this sample
    are shown in black, while others are shown in gray.
    The solid red line marks the adopted BD-limited, mass-selected
    boundary at 0.04\,$M_\odot$ on the dwarf track, drawn parallel to
    the extinction vector.
    The dotted red line, passing through 0.08\,$M_\odot$ and drawn
    parallel to the extinction vector, indicates the canonical
    stellar/brown-dwarf boundary used for robustness
    checks.}

\label{fig:cc_nircams}
\end{center}
\end{figure}

\begin{figure}[h]
  \begin{center}
    \includegraphics[scale=0.7]{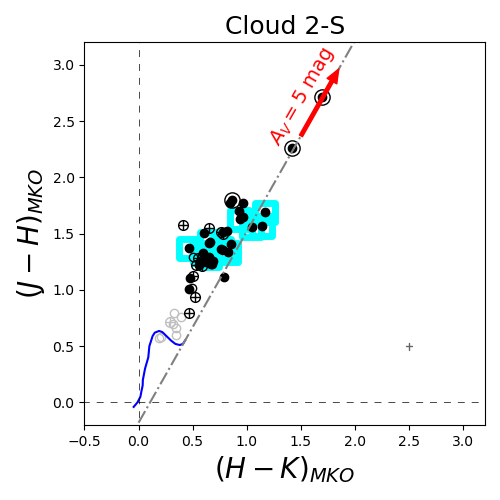}
    \includegraphics[scale=0.7]{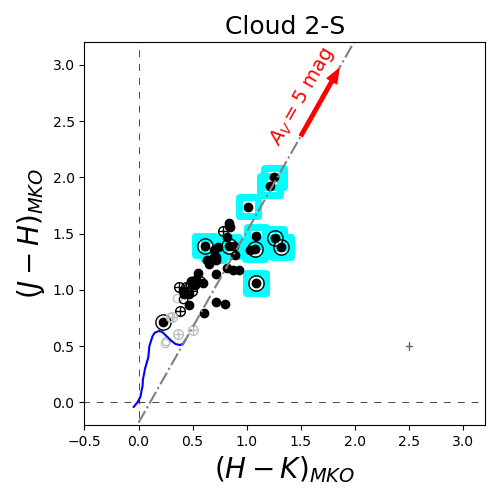}
    \caption{JHK color--color diagram from a previous study
      \citep{Yasui2009}, based on data obtained with Subaru/MOIRCS.
      The sources shown here are those matched with objects for which
      the MIR SED slopes ($\alpha$) are derived in this study.
      The symbols are the same as in Figure~\ref{fig:CM_MIRI_NIRCam},
      according to the derived $\alpha$.
      Sources within the $A_V$-limited sample, as defined in
      Figure~\ref{fig:cc_nircams}, are shown in black, while others
      are shown in gray.
      Typical photometric uncertainties for these sources are
      indicated by thick gray error bars in the lower-right corner.
      The blue curves indicate the locus of points corresponding to
      unreddened main-sequence stars in the MKO filter system.
      The gray dot--dashed lines, which pass through the point
      corresponding to spectral type M6 on the track and are drawn
      parallel to the extinction vector, indicate the boundary between
      disk-excess and disk-free sources.}
 \label{fig:cc_mcs}
\end{center}
\end{figure}

\begin{figure}
\begin{center}
 \includegraphics[scale=0.7]{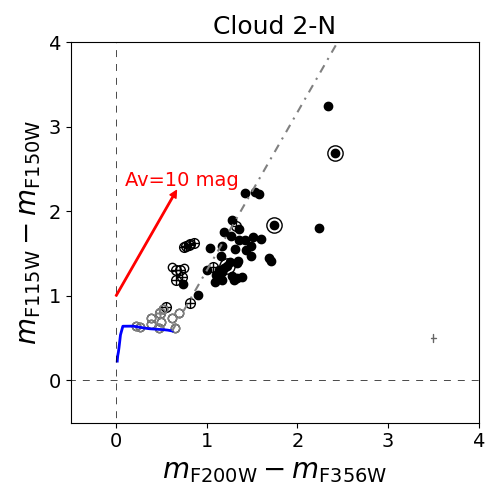}
 \includegraphics[scale=0.7]{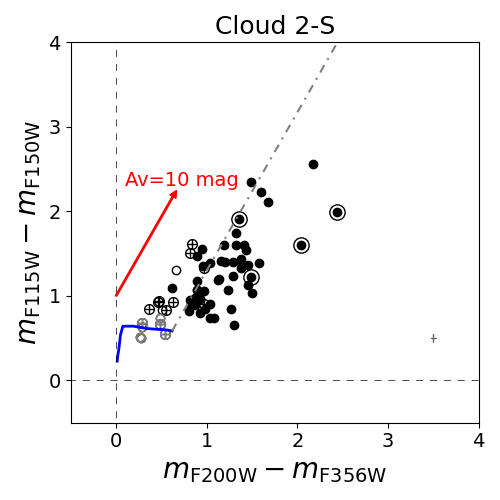}

 \caption{F356W-excess diagnostic diagram, 
   $(m_{\rm F200W} - m_{\rm F356W})$ vs. $(m_{\rm F115W} - m_{\rm
     F150W})$, for the Cloud 2 clusters.
   Sources in the cluster regions but not detected in F770W band are
   shown as large open squares.
   Typical photometric uncertainties are indicated by thick gray error
   bars in the lower-right corner.
   The small circles show the sources detected in all F356W, F444W,
   and F770W bands, with the same symbols as in
   Figure~\ref{fig:CM_MIRI_NIRCam}, based on MIR SED slope ($\alpha$).
   Sources within the $A_V$-limited sample, as defined in
   Figure~\ref{fig:cc_nircams}, are shown in black, while others are
   shown in gray.
   {The blue curves represent the dwarf tracks.}
   The gray dot-dashed line marks the boundary between disk-excess and
   disk-free sources, defined to be parallel to the reddening vector
   and passing through the 0.1 $M_\odot$ point on the {dwarf track.}}
  \label{fig:CC_NIRCam_F356W}
\end{center}
\end{figure}

\begin{figure}
\begin{center}
  \includegraphics[scale=0.7]{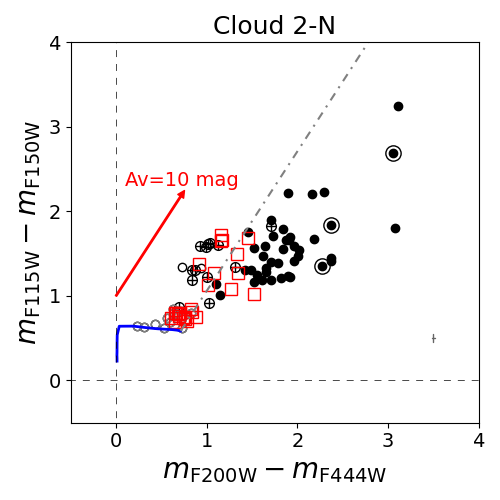}
 \includegraphics[scale=0.7]{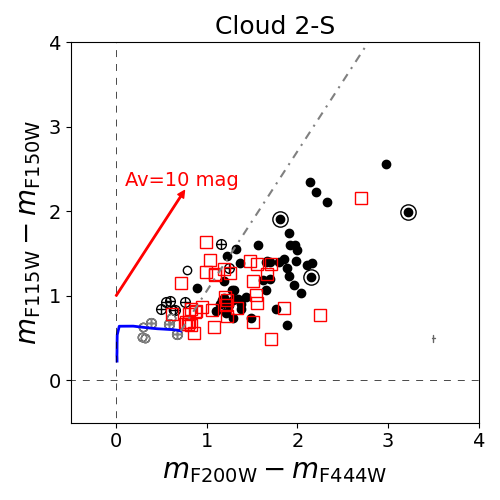}
 \caption{F444W-excess diagnostic diagram, $(m_{\rm F200W} - m_{\rm
     F444W})$ vs. $(m_{\rm F115W} - m_{\rm F150W})$, for the Cloud 2
   clusters.
   The circles represent the sources detected in all F356W, F444W, and
   F770W bands, with the same symbols as in
   Figure~\ref{fig:CM_MIRI_NIRCam}, based on MIR SED slope ($\alpha$).
   Typical photometric uncertainties for these sources are indicated
   by thick gray error bars in the lower-right corner.
   Sources within the $A_V$-limited sample, as defined in
   Figure~\ref{fig:cc_nircams}, are shown in black, while others are
   shown in gray.
   Sources for which $\alpha$ could not be derived, primarily due to
   non-detection in the F770W band, are indicated by red open squares.
   The blue curves represent the dwarf tracks.
   The gray dot-dashed line marks the boundary between disk-excess and
   disk-free sources, defined to be parallel to the reddening vector
   and passing through the 0.1 $M_\odot$ point on the {dwarf
     track.}}
 \label{fig:CC_NIRCam_F444W}
\end{center}
\end{figure}

\begin{figure}
\begin{center}
  \includegraphics[scale=0.7]{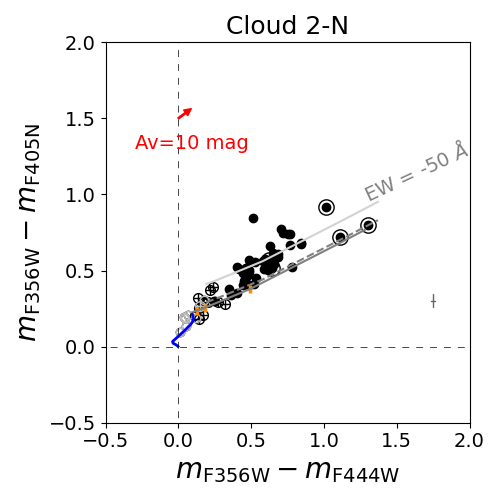}
 \includegraphics[scale=0.7]{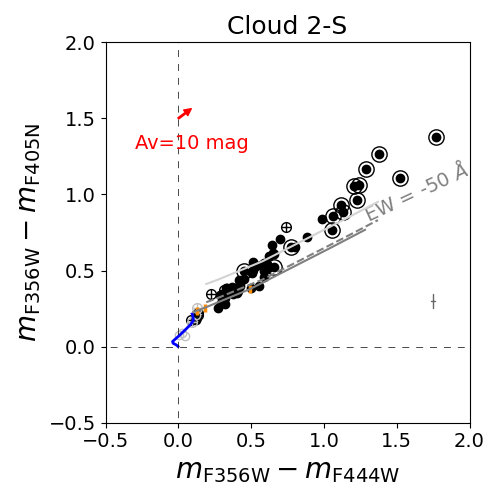}
 \caption{$(m_{\rm F356W} - m_{\rm F444W})$ vs. $(m_{\rm F356W} -
   m_{\rm F405N})$ color--color diagram for the Cloud 2-N and 2-S
   clusters.
   The symbols are the same as in Figure~\ref{fig:CM_MIRI_NIRCam},
   based on MIR SED slope ($\alpha$).
   Typical photometric uncertainties are indicated by thick gray error
   bars in the lower-right of each panel.
   Sources within the $A_V$-limited sample, as defined in
   Figure~\ref{fig:cc_nircams}, are shown in black, while others are
   shown in gray.
   Class I sources outside the $A_V$-limited sample are also included,
   as they are likely cluster members despite lacking $A_V$ estimates.
   The blue curves represent the dwarf tracks.
   Dark gray curves intersect the dwarf track at a stellar mass of 0.1
   $M_\odot$ and trace the displacement associated with increasing
   $\alpha$.
   The boundaries between the classes---Class III and EV ($\alpha =
   -2.56$), EV and Class II ($\alpha = -1.8$), and Class II and Class
   I ($\alpha = 0$)---are marked by orange tips for clarity.
   The dashed gray curve shows the $\alpha$-induced displacement
   shifted according to the reddening vector corresponding to $A_V=
   10$ mag. The light gray curve is obtained by further shifting this
   reddening-adjusted $\alpha$-displacement along the direction
   corresponding to Br$\alpha$ emission with an equivalent width of
   $-$50 \AA.}
 \label{fig:CC_f405n_f356w_f444w}
\end{center}
\end{figure}

\begin{figure}
  \begin{center}
  \includegraphics[scale=0.55]{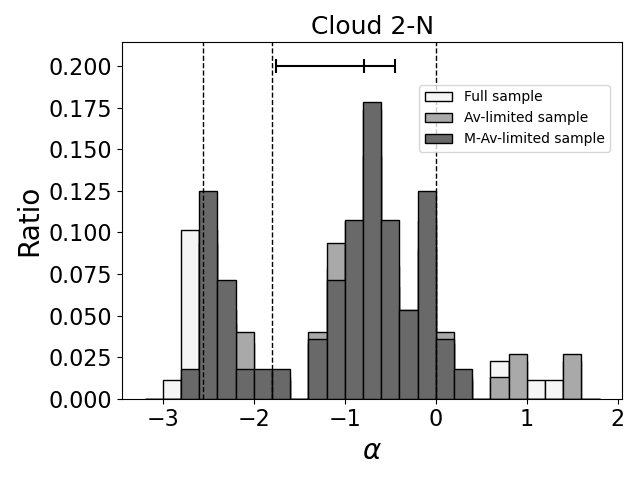}
  \includegraphics[scale=0.55]{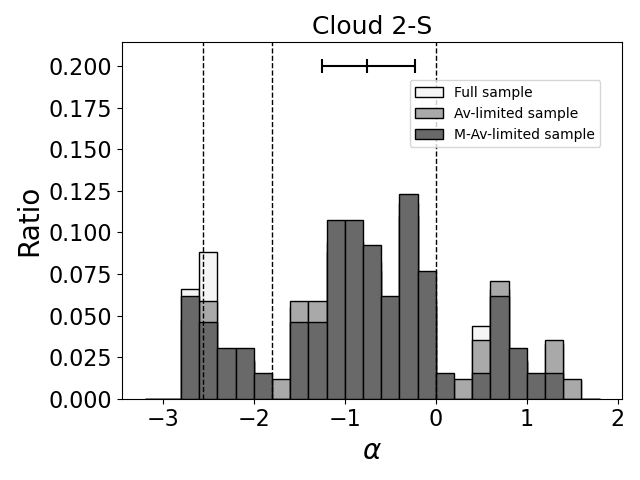}
 \caption{Distribution of MIR SED slope ($\alpha$) of sources for the
   Cloud 2-N cluster (left) and Cloud 2-S cluster (right).
   The distributions for sources within the full sample, $A_V$-limited
   sample, and the Mass-$A_V$-limited sample are shown with white,
   light gray and dark gray histograms, respectively.}
 \label{fig:sed_dist}
\end{center}
\end{figure}

\begin{figure}
\begin{center}
 \includegraphics[scale=0.7]{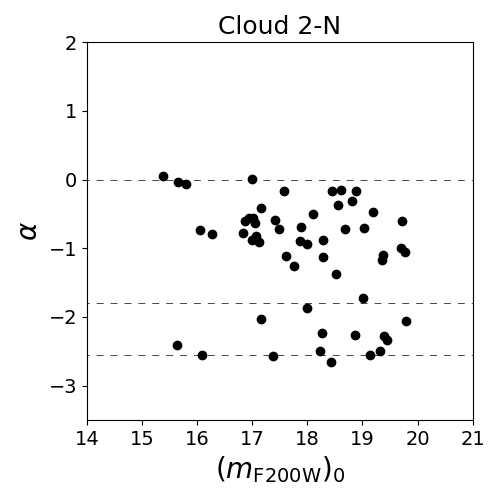}
 \includegraphics[scale=0.7]{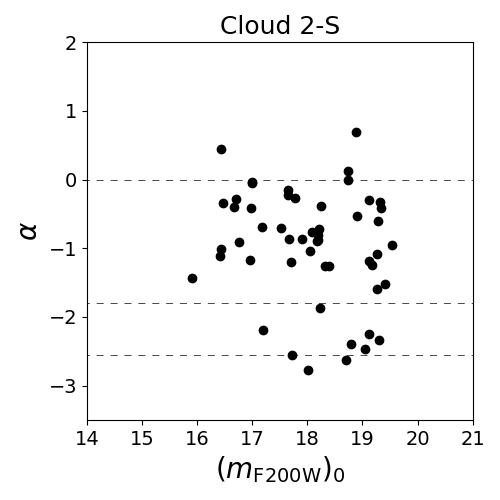}
  \caption{Scatter plot of {intrinsic $m_{\rm F200W}$ ($(m_{\rm
        F200W})_0$)} vs. MIR SED slope ($\alpha$) for sources that are
    included in the Mass-$A_V$-limited sample defined in
    Section~\ref{sec:disk_bd}.
    The horizontal dashed lines at $\alpha = 0$, $-$1.8, and $-$2.56
    indicate the boundaries used to classify sources into Class I,
    Class II, EV, and Class III, respectively.}
 \label{fig:sed_f200w}
\end{center}
\end{figure}

\begin{figure}
\begin{center}
  \includegraphics[scale=0.35]{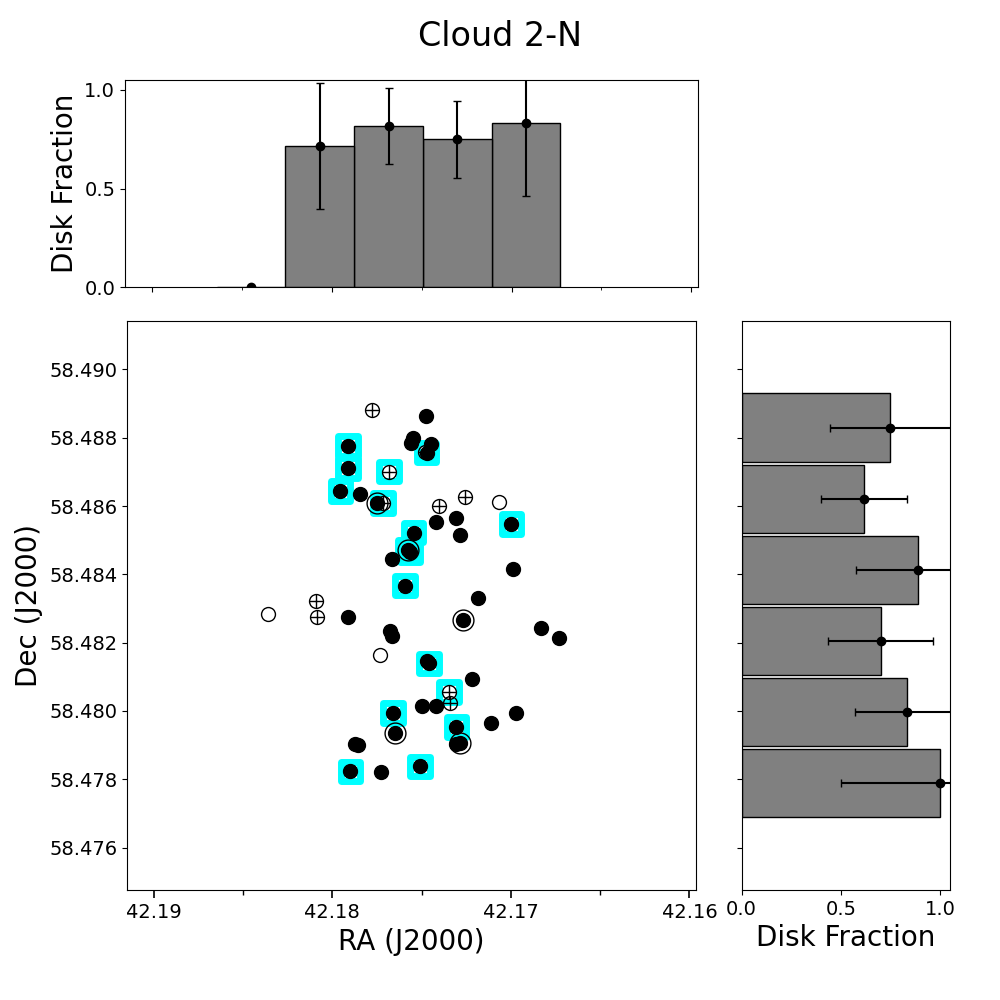}
  \includegraphics[scale=0.35]{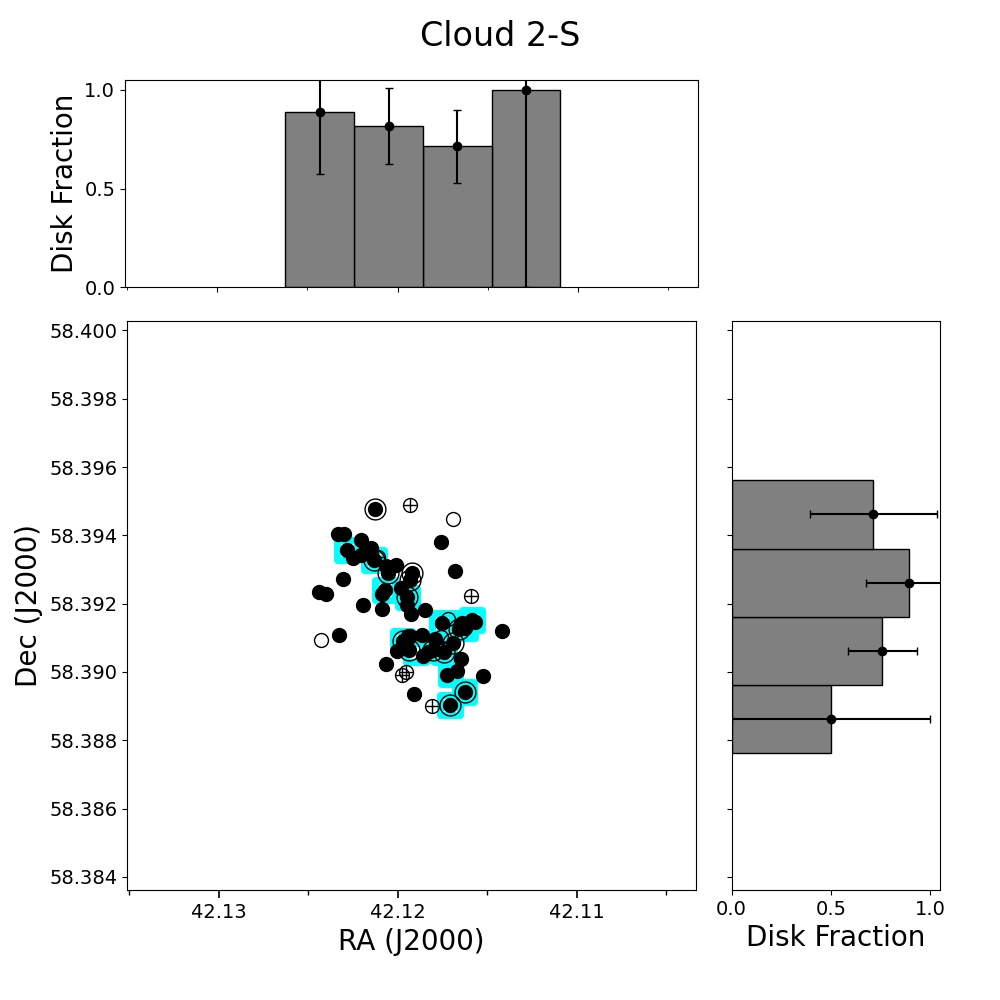}
  \caption{Spatial distribution of disk-classified sources and
    optically thick disk fractions based on MIR SED slope
    ($\alpha$). The figure consists of two sets of panels,
    corresponding to the Cloud 2-N cluster (left) and the Cloud 2-S
    cluster (right). In each set, the bottom-left panel shows the
    positions of sources within the cluster region. Only sources
    belonging to the Mass–$A_V$–limited sample and Class I sources are
    plotted to provide a more complete view of the disk-bearing
    population. Both groups are shown with the same symbols as in
    Figure~\ref{fig:CM_MIRI_NIRCam}.
    Sources exhibiting Br$\alpha$ excess are additionally marked with
    cyan rectangles. The top-left and bottom-right panels in each set
    show the spatial distributions of optically thick disk fractions
    derived from the $\alpha$-based classification, binned in two
    directions: east–west (top-left) and north–south
    (bottom-right). The histograms represent the disk fractions, and
    the error bars indicate Poisson uncertainties.}
 \label{fig:distribution}
\end{center}
\end{figure}

\begin{figure}
\begin{center}
  \includegraphics[scale=0.7]{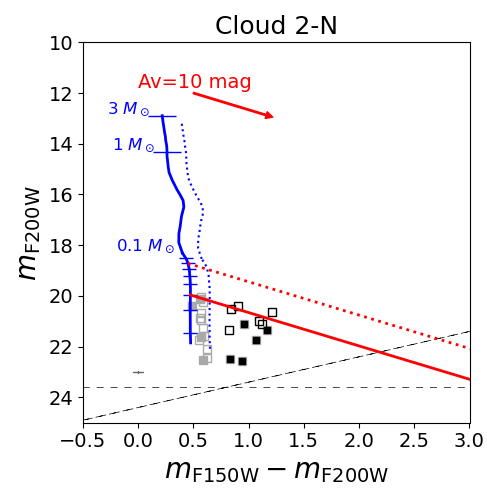}
 \includegraphics[scale=0.7]{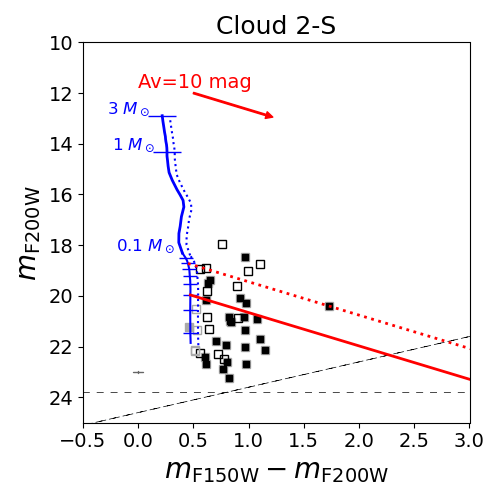}
  \caption{Same color--magnitude diagram as in
    Figure~\ref{fig:cc_nircams}, $(m_{\rm F115W} - m_{\rm F200W})$
    vs. $m_{\rm F200W}$, but for sources that are not detected in the
    F770W band and detected up to the F444W band.
    Squares represent sources for which $\alpha$ could not be
    derived, primarily due to non-detection of the F770W band.
    Typical photometric uncertainties for these sources are indicated
    by thick gray error bars in the lower-left corner.
    Among these, filled squares indicate objects identified as having
    disks based on F444W excess, while open squares indicate objects
    without disks.
    Sources within $A_V$-limited sample are shown in black, while
    others are shown in gray.
    The mass-selection boundaries are the same as in
    Figure~\ref{fig:cc_nircams}: the solid red line marks the
    operational BD-limited boundary at 0.04 $M_\odot$, and the dotted
    red line marks the canonical stellar/brown-dwarf boundary at 0.08
    $M_\odot$.}
  \label{fig:cc_nircam_bd}
\end{center}
\end{figure}

\clearpage
\appendix

\section{Observed MIR SEDs for sources in the Cloud 2 clusters}
\label{sec_appendix:SED}

For sources detected in the F770W wide-band filter within the cluster
region, the photometric results from the F115W to F770W wide-band
filters (F115W, F150W, F200W, F356W, F444W, and F770W) are shown as
black circles, while the photometric results from the F405N
narrow-band filter are shown as open triangles.
The MIR SED slope ($\alpha$), derived in Section~\ref{sec:sed}, is
indicated by black lines.
The results for the Cloud 2-N clusters are presented in
Figure~\ref{fig:SED_CL2N}, and those for the Cloud 2-S clusters are
shown in Figure~\ref{fig:SED_CL2S}.

\section{Comparison with Results from Previous NIR Observations}
\label{sec_appendix:NIRobs}

To examine the correspondence between the presence of disks around
individual objects, as determined using NIR data up to 2 $\mu$m, and
the results of this study, we use the Subaru/MOIRCS data employed by
\citet{Yasui2009}.
These data are suitable for direct comparison with the F770W band data
used to derive $\alpha$ in this paper, as they have similar mass
limits and spatial resolution (mass limit of $\sim$0.1 $M_\odot$ and
spatial resolution of approximately 0.3 arcsec).
Additionally, the JWST filter system currently lacks fully established
tools for accurately determining disk excess and disk fraction in the
NIR bands. In contrast, detailed observations have been conducted
under the MKO filter system adopted by Subaru/MOIRCS, and methods for
estimating disk excess are well-established. This is the reason for
using data from past literature rather than JWST data.

To determine the color excess, the following tools are
necessary. Specifically, precise photometry results for objects with
known spectral types \citep{Leggett2006}, discussion of color changes in
different filter systems \citep{Hewett2006}, and the reddening law
\citep{Nishiyama2006ApJ}.
By integrating these, the dwarf track, CTTS locus, and reddening law
in the MKO system were examined \citep{Yasui2008}.
When deriving the disk fraction from data up to 2 $\mu$m in the NIR,
the H-K vs. J-H color–color diagram (hereafter, JHK color-color
diagram) is generally used (Figure~\ref{fig:cc_mcs}).
In this diagram, a line (dot-dashed lines) passing through the point
corresponding to the M6 type star on the dwarf star track (blue curve)
and parallel to the reddening vector is set as the boundary for
determining the presence or absence of a disk.
The lower right side of the boundary line is considered the disk
excess region, while the upper left side is considered the disk-free
region.
The fraction of objects located in the disk excess region
out of the total is defined as the NIR disk fraction.

We compared the coordinates of the photometry data from
\citet{Yasui2009} with those of the photometry data obtained in this
study and performed coordinate matching within a 0.5 arcsec radius
using Astropy.
For the matched data, we plotted them on the NIR color--color diagram
(Figure~\ref{fig:cc_mcs}), color-coded according to the disk
classification based on $\alpha$ from this study.

\section{Protoplanetary disks around brown dwarfs} \label{sec_appendix:cc_dwarf}

The protoplanetary disks of brown dwarfs have been the subject of
investigation and discussion since approximately 2005, utilizing the
Spitzer telescope and, in many cases, employing wavelengths up to 5.8
$\mu$m (IRAC band 3 in the Spitzer/IRAC).
In some cases, the presence or absence of a disk has been discussed
based on spectroscopic observations, but here we discuss previous
research that has also dealt with imaging observations in order to
discuss the existence of a brown dwarf disk based on imaging data.
As previously discussed in Section~\ref{sec:phot}, the sensitivity and
spatial resolution of Spitzer's observations of nearby star-forming
regions are approximately equivalent to those of JWST's observations
of outer Galaxy objects.
Therefore, the findings and discussions from previous studies are
directly applicable to the analysis of the targets presented in this
work.

In previous studies, brown dwarf objects exhibiting significant color
excesses at 5.8 $\mu$m and 4.5 $\mu$m on both axes of the ([4.5] -
[5.8]) vs. ([3.6] - [4.5]) color–color diagram (upper left panel of
Figure~\ref{fig:cc_BD}) are generally classified as having disks. The
classification criteria adopted by \citet{Luhman2005}, shown as dashed
lines in the diagram, define disk-bearing objects as those located to
the upper right of the boundary: $[4.5] - [5.8] \ge 0.16$ mag and
$[3.6] - [4.5] \ge 0.2$ mag.
In Figure~\ref{fig:cc_BD}, data from nearby star-forming regions
previously discussed in the context of brown dwarf disks and observed
through Spitzer/IRAC imaging are plotted: Chamaeleon I (cross;
\citet{Luhman2005}, IC 348 (dot; \citet{Luhman2005}), Taurus (circle
for \citealt{Monin2010}, triangle for \citealt{Guieu2007}), and Upper
Sco (plus; \citealt{Riaz2009}). Only sources with available data in
all three bands (3.6, 4.5, and 5.8 $\mu$m) are shown in
Figure~\ref{fig:cc_BD}.
The brown dwarfs included in Figure~\ref{fig:cc_BD} were confirmed as
substellar through spectroscopic observations in the original studies.

Objects identified as possessing disks according to the original
classifications in the respective literature are indicated in red,
while those classified as diskless are shown in black. It is
demonstrated that the disk classifications based on the
\citet{Luhman2005} criteria are generally consistent with those
reported {in the literature \citep{{Monin2010}, {Guieu2007},
    {Riaz2009}}}.
The only exception is one object in Taurus reported by
\citealt{Guieu2007}, {which is shown with a gray filled triangle
  in Figure~\ref{fig:cc_BD}.}  Although this object is classified as
having a disk based on the diagram here, it was reported as diskless
in the original study.
For reference, {the dwarf track is shown as a blue curve, and the
  reddening vector for $A_V = 10$ mag is shown with a red arrow.}
From this figure, it can be seen that objects with disks tend to show
significant color excesses along both axes.

We have independently examined the behavior of these brown dwarf
samples using supplementary diagnostics for nearby star-forming
regions (the corresponding diagrams are not shown here) and find no
obvious evidence for a mass-dependent variation in disk fraction or
for an extremely high-extinction subpopulation.

Using the brown dwarf samples plotted in Figure~\ref{fig:cc_BD} from
these nearby star-forming regions, we investigated whether disk
presence can be diagnosed using photometric data at wavelengths
shorter than 5 $\mu$m. Several color--color diagrams combining NIR JHK
bands and IRAC bands were examined to determine an effective and
consistent criterion. In order to follow the conventional approach of
previous brown-dwarf disk studies, we adopted the dwarf track as the
basis for defining the disk/no-disk boundary.

Among the diagrams considered, we use the $K-{\rm [4.5]}$ vs. $J-H$
plot (right panel of Figure~\ref{fig:cc_BD}) as our primary
diagnostic, because it combines two bands widely separated in
wavelength and thus enhances sensitivity to infrared excess from
disks.
Typical photometric uncertainties are shown for reference, indicating
that, except for sources lying close to the boundary, the separation
between disk-bearing and disk-free objects is broadly comparable
to---or larger than---the representative errors.

In this diagram, we define the disk/no-disk boundary by drawing a line
parallel to the reddening vector and passing through the M6 point
(corresponding to $\simeq$0.1 $M_\odot$) on the dwarf track. Most
disk-bearing objects lie on the excess side of this boundary, while a
small number of sources appear as outliers on either side, likely
reflecting measurement uncertainties and/or modest extinction effects
at J and H. Overall, this criterion provides a practical and
consistent basis for disk classification using bands shorter than 5
$\mu$m.

Figure~\ref{fig:cc_BD} presents two representative diagnostics: (1)
the original disk-identification diagram of \citet{Luhman2005}, and
(2) the K–[4.5] versus J–H diagram illustrating our adopted
approach. Therefore, in this study, we adopt this method to determine
the disk fraction of brown dwarfs in the Cloud 2 clusters (see
Section~\ref{sec:disk_bd}).

\begin{figure}
  \begin{center}
    \includegraphics[scale=0.55]{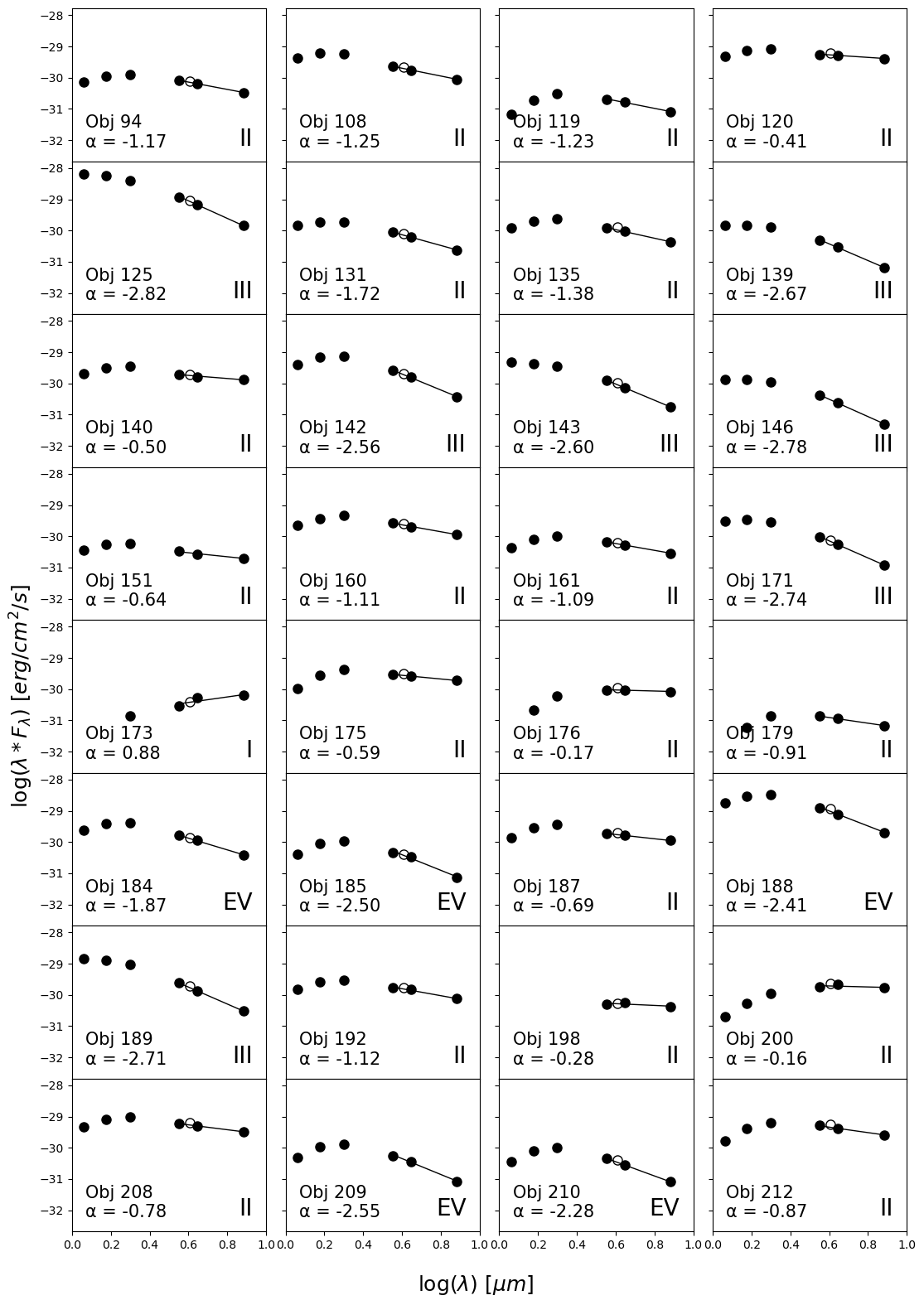}
    \caption{Observed SEDs of sources in the Cloud 2-N cluster
      detected in the F770W band. Black lines indicate the results of
      fits used to determine the MIR SED slopes ($\alpha$). Filled
      circles represent fluxes detected in wide-band filters, while
      open circles indicate fluxes detected in the F405N narrow-band
      filter, which covers the Br$\alpha$ emission line.}
\label{fig:SED_CL2N}
\end{center}
\end{figure}

\setcounter{figure}{14}
\begin{figure}
  \begin{center}
    \includegraphics[scale=0.55]{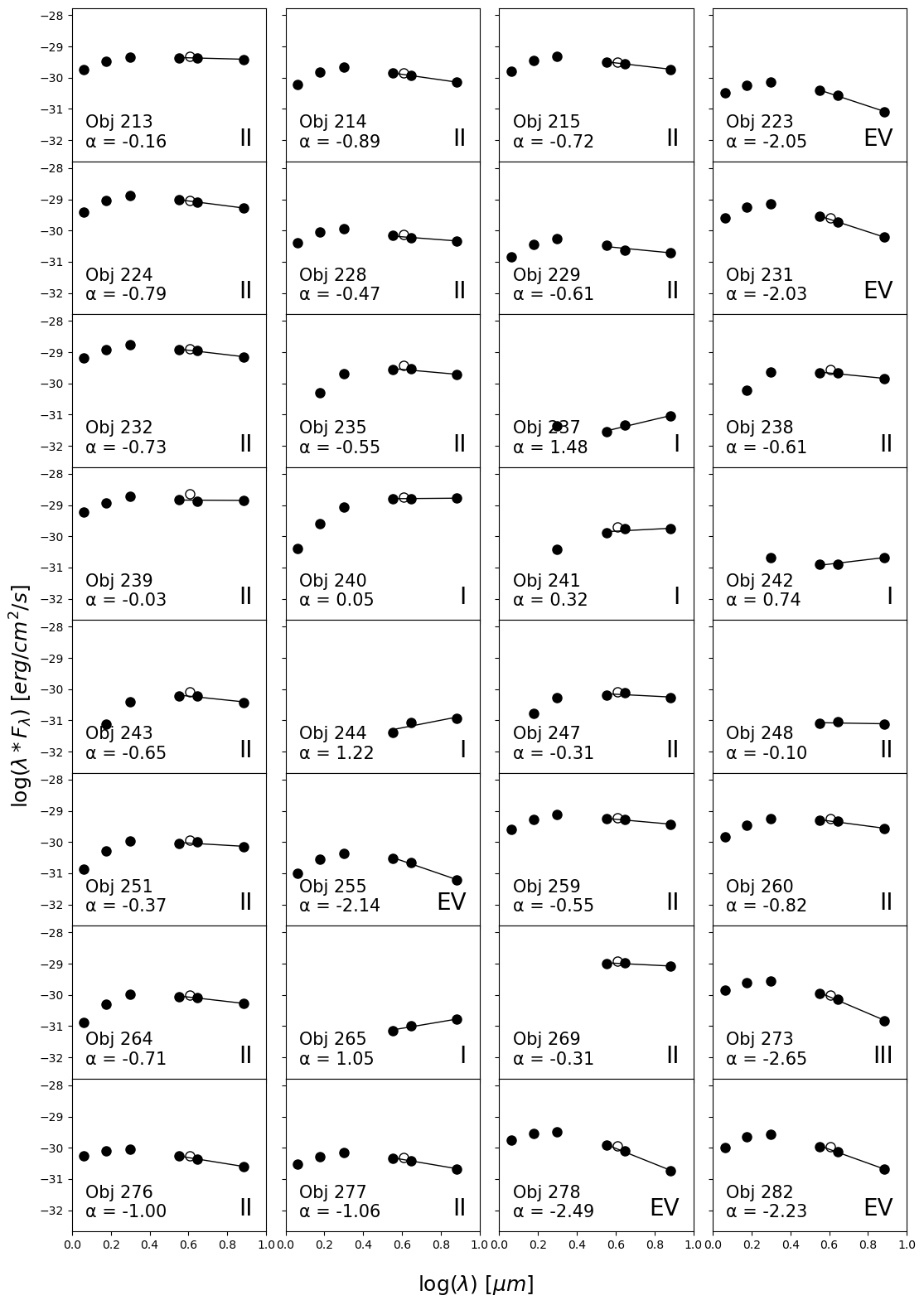}
 \caption{(Continued) Observed SEDs of sources in the Cloud 2-N cluster.}
\label{fig:SED_CL2N}
\end{center}
\end{figure}

\setcounter{figure}{14}
\begin{figure}
  \begin{center}
    \includegraphics[scale=0.55]{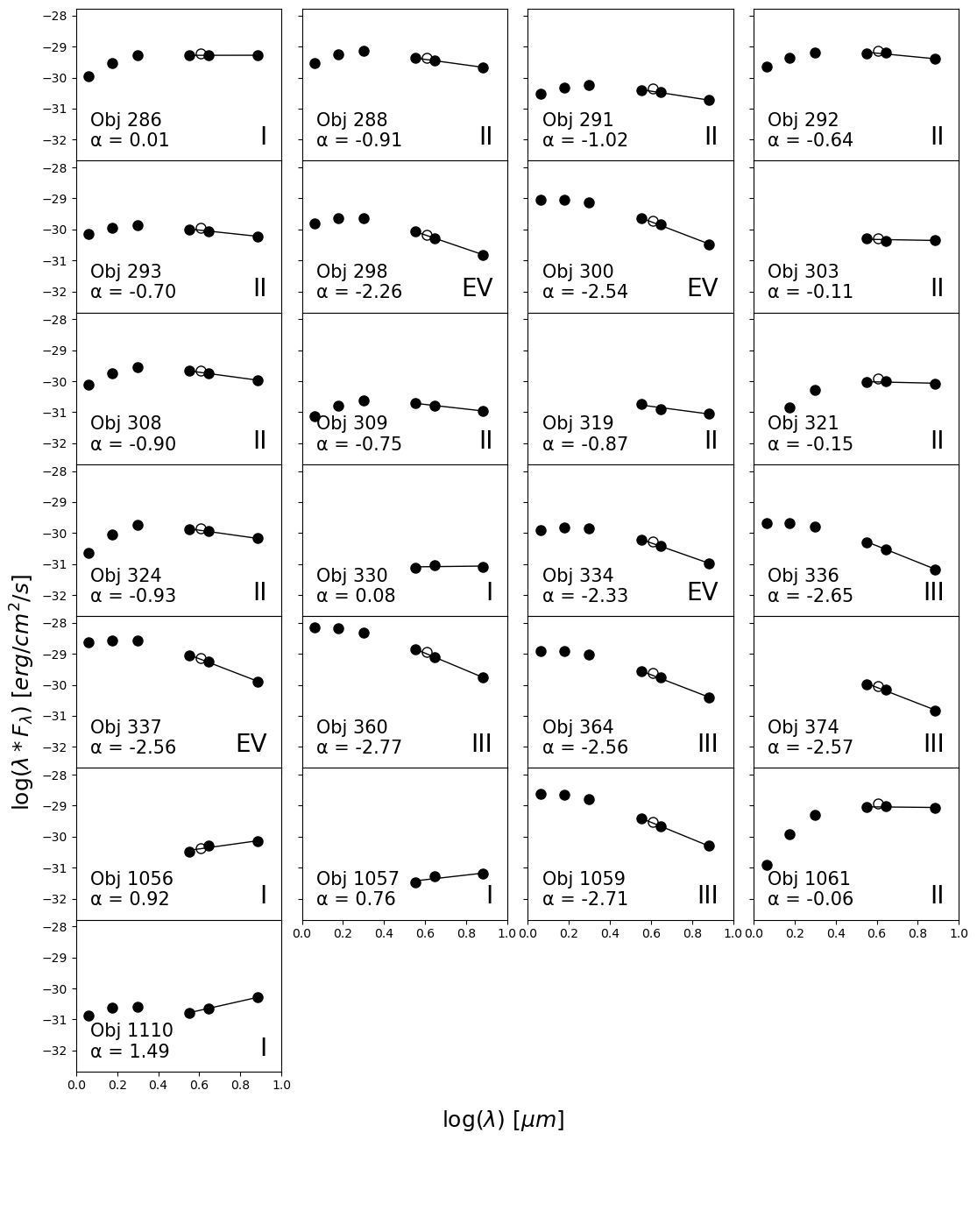}
 \caption{(Continued) Observed SEDs of sources in the Cloud 2-N cluster.}
\label{fig:SED_CL2N}
\end{center}
\end{figure}


\begin{figure}[h]
  \begin{center}
    \includegraphics[scale=0.55]{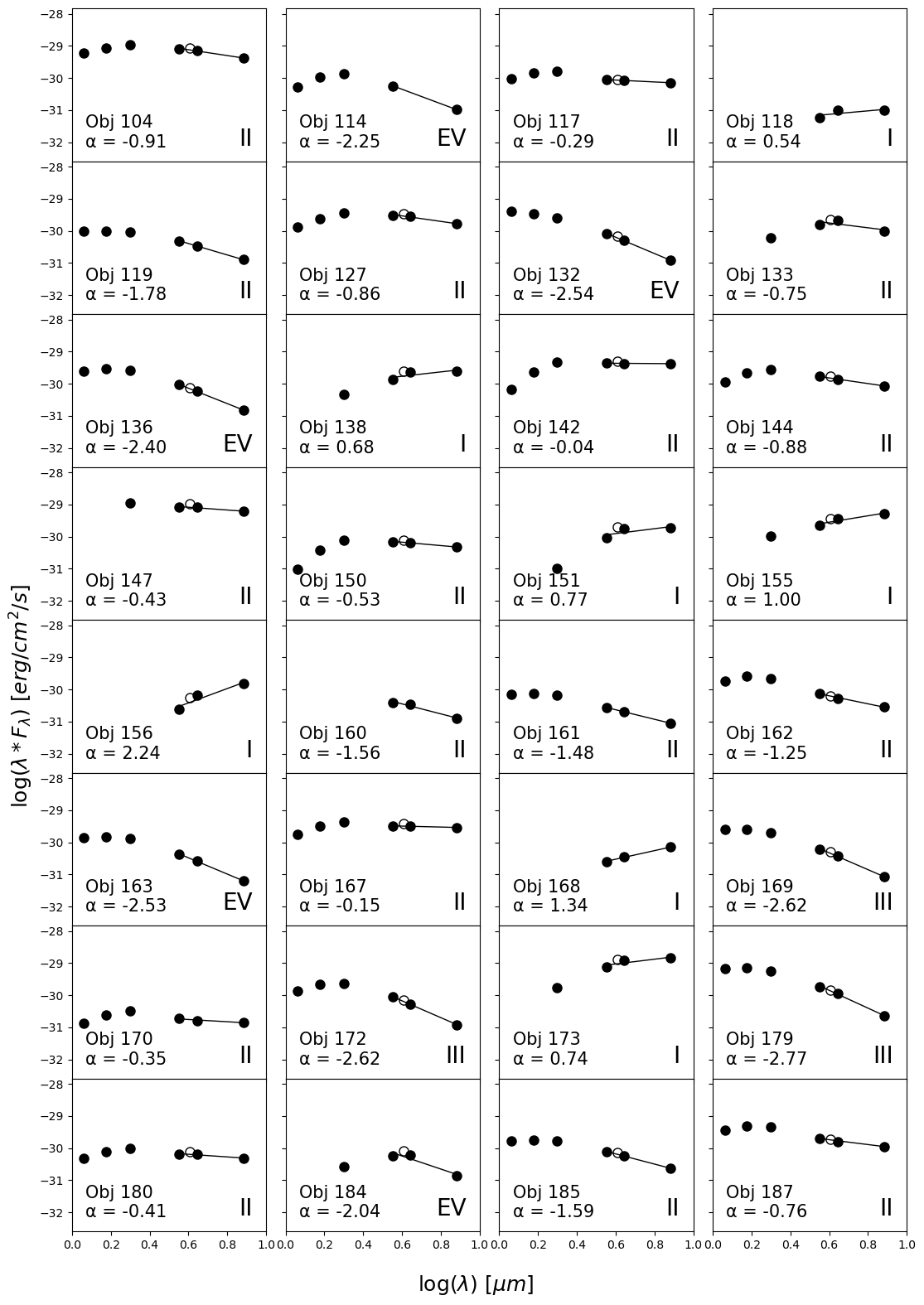}
    \caption{The same as Figure~\ref{fig:SED_CL2N}, but for the Cloud
      2-S cluster.}
\label{fig:SED_CL2S}
\end{center}
\end{figure}

\setcounter{figure}{15}
\begin{figure}[h]
  \begin{center}
    \includegraphics[scale=0.55]{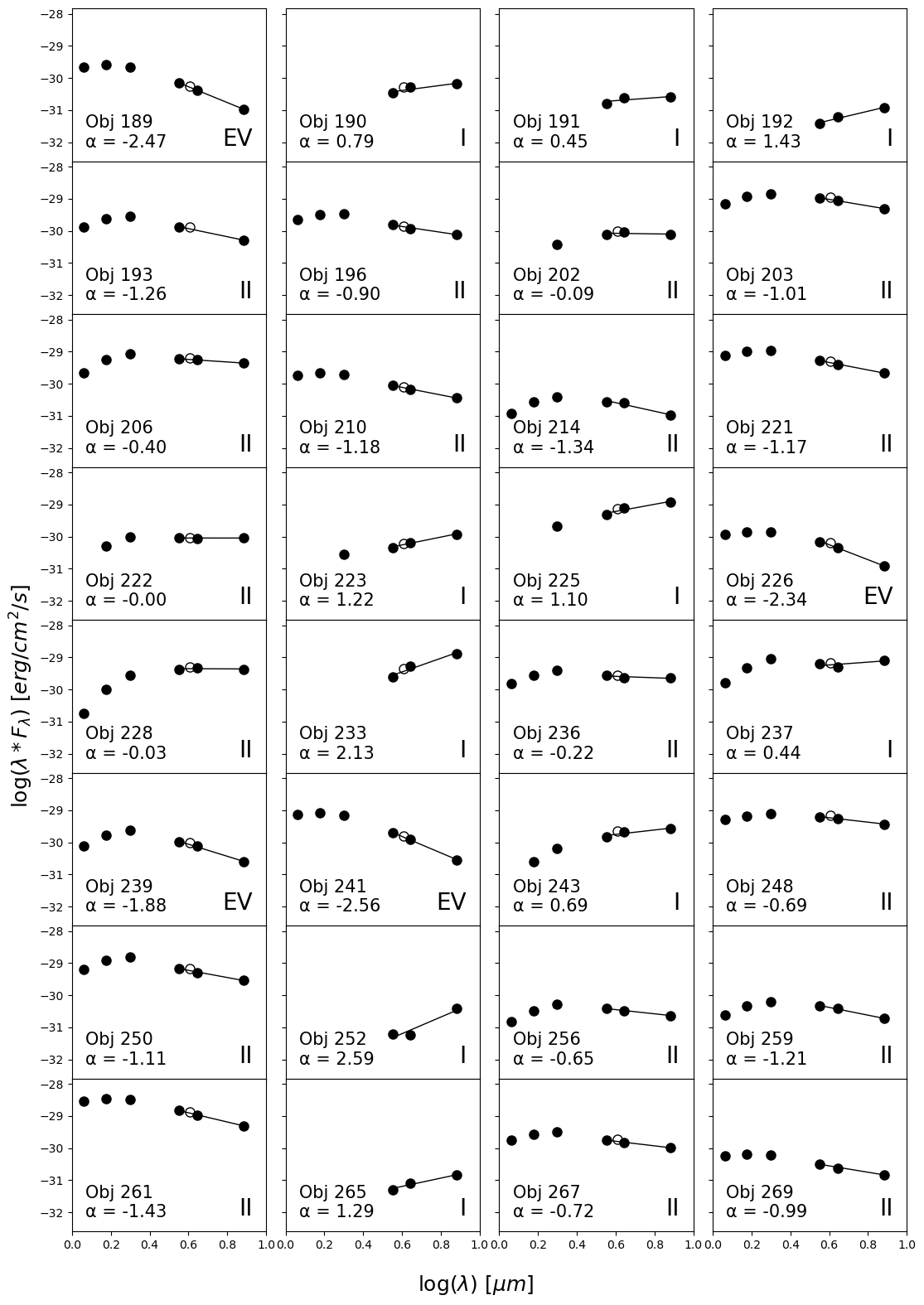}
 \caption{(Continued) Observed SEDs for sources in the Cloud 2-S cluster region.}
\end{center}
\end{figure}

\setcounter{figure}{15}
\begin{figure}[h]
  \begin{center}
    \includegraphics[scale=0.55]{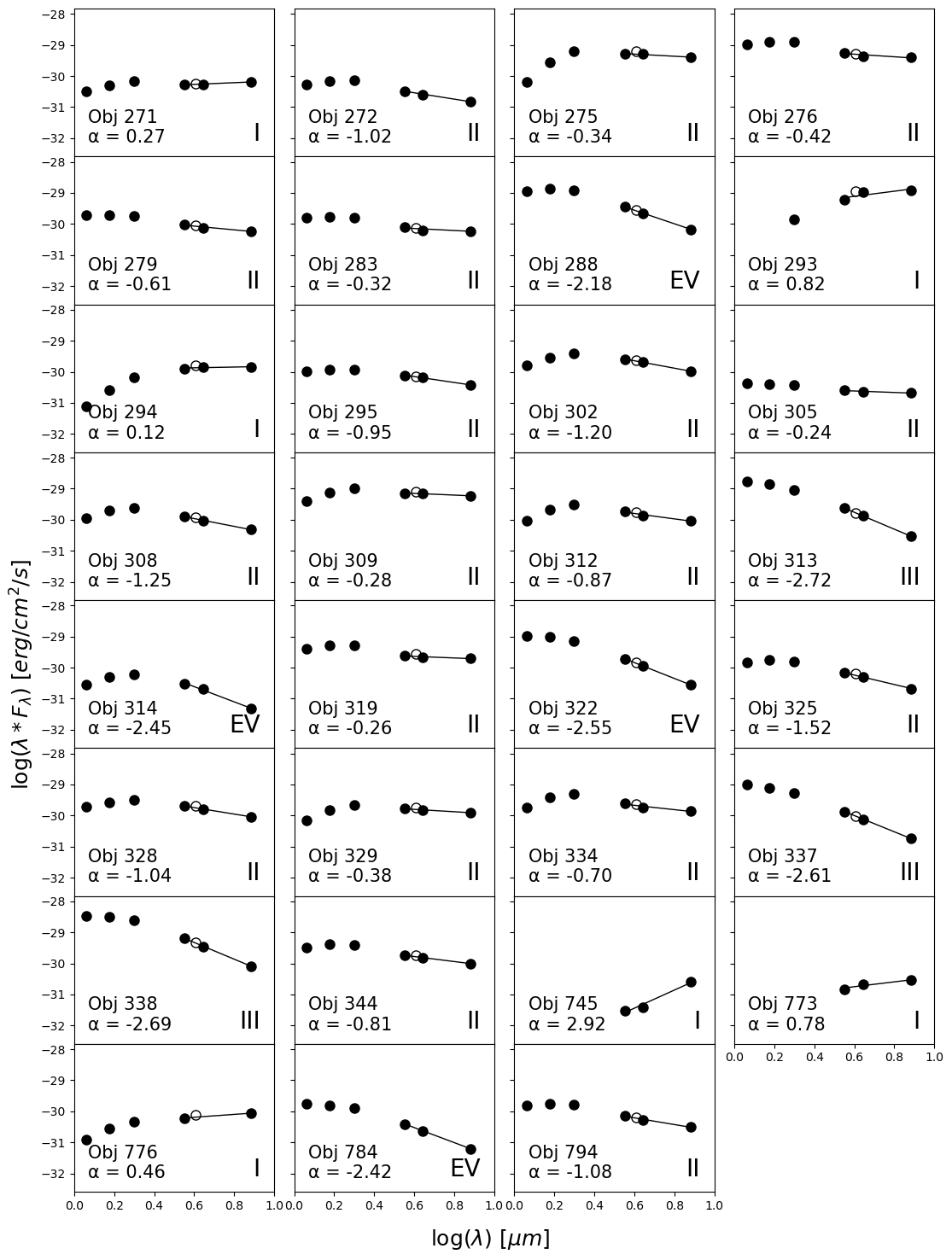}
 \caption{(Continued) Observed SEDs for sources in the Cloud 2-S cluster region.}
\end{center}
\end{figure}

\begin{figure}
\begin{center} 
 \includegraphics[scale=0.55]{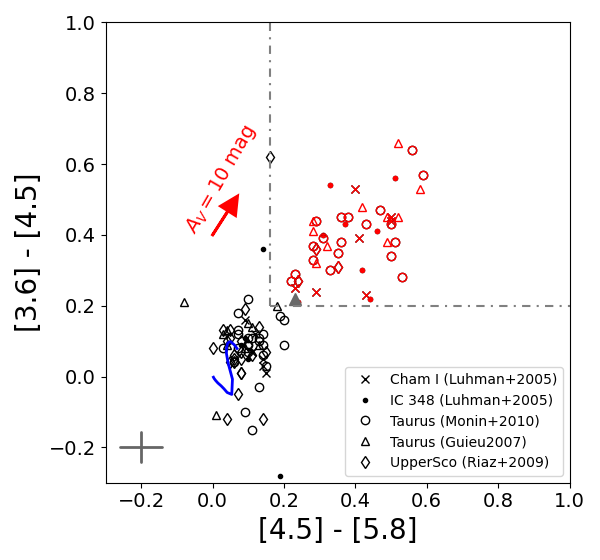}
 \includegraphics[scale=0.55]{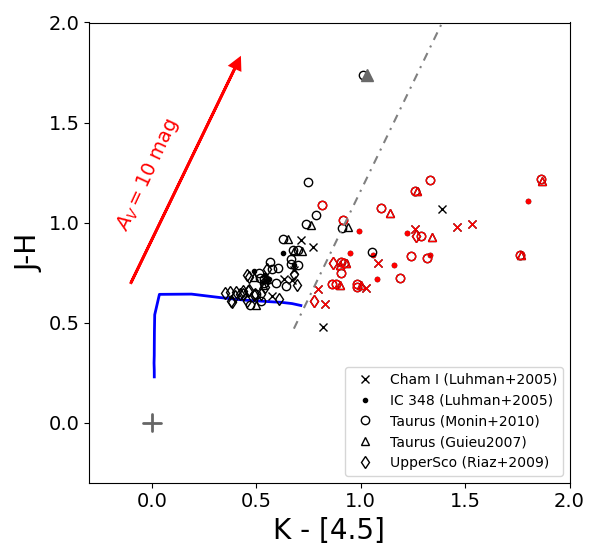}
 \caption{Color–color diagrams for brown dwarfs (BDs), using bands up
   to 4--5 $\mu$m, based on data from previous studies.
   The data are taken from literature on nearby star-forming
   regions: Chamaeleon I (crosses; \citealt{Luhman2005}), IC 348
   (dots; \citealt{Luhman2005}), Taurus (circles from
   \citealt{Monin2010} and triangles from \citealt{Guieu2007}), and
   Upper Scorpius (diamonds; \citealt{Riaz2009}).
   {The left panel uses bands up to 5 $\mu$m and corresponds to
     Luhman et al. (2005), where the region above the gray dot-dashed
     line indicates objects with disks. The right panel uses bands up
     to 4 $\mu$m and demonstrates our proposed approach for disk
     classification.}
   {Typical photometric uncertainties are indicated by thick gray
     error bars in the lower-right corners of both panels.}
   {The blue curves in both two panels represent main sequence
     dwarf track.}
   Red arrows indicate the reddening vector for an extinction of
   $A_V=10$ mag. In {the right panels}, the gray dot-dashed line
   passes through the 0.1 $M_\odot$ point and is drawn parallel to the
   reddening vector, serving as a boundary for disk classification.}
 \label{fig:cc_BD}
\end{center}
\end{figure}

\end{document}